\documentclass[twocolumn]{article}
\pdfoutput=1
\usepackage[margin=0.8in]{geometry}
\usepackage[numbers, sort&compress]{natbib}
\usepackage{graphicx}
\usepackage{url}
\usepackage{amsmath}
\usepackage{amssymb}
\usepackage{siunitx}
\DeclareSIUnit\Molar{M}
\usepackage{booktabs}
\usepackage{authblk}

\makeatletter
\renewcommand{\maketitle}{\bgroup\setlength{\parindent}{0pt}
  \textbf{\LARGE \@title}
\begin{flushleft}
  \@author
\end{flushleft}\egroup
\hfill \break
\hfill \break
\hfill \break
\hfill \break
\hfill \break
}
\makeatother

\title{How sticky are our proteins? Quantifying hydrophobicity of the human proteome}

\author[1*]{Juami Hermine Mariama van Gils}
\author[1]{Dea Gogishvili}
\author[1]{Jan van Eck}
\author[1]{Robbin Bouwmeester}
\author[1]{Erik van Dijk}
\author[1*]{Sanne Abeln}
\affil[1]{Center for Integrative Bioinformatics (IBIVU), Computer Science Department, VU University, Amsterdam, 1081 HV, The Netherlands}
\affil[*]{Corresponding authors; s.abeln@vu.nl, j.h.m.van.gils@vu.nl}

\date{July 2021}

\begin{document}
\begin{titlepage}
\maketitle

\begin{abstract}
Proteins tend to bury hydrophobic residues inside their core during the folding process to provide stability to the protein structure and to prevent aggregation. Nevertheless, proteins do expose some 'sticky' hydrophobic residues to the solvent. These residues can play an important functional role, for example in protein-protein and membrane interactions.
Here, we investigate how hydrophobic protein surfaces are by providing three measures for surface hydrophobicity: the total hydrophobic surface area, the relative hydrophobic surface area, and - using our MolPatch method - the largest hydrophobic patch.
Secondly, we analyse how difficult it is to predict these measures from sequence: by adapting solvent accessibility predictions from NetSurfP2.0, we obtain well-performing prediction methods for the THSA and RHSA, while predicting LHP is more difficult.
Finally, we analyse implications of exposed hydrophobic surfaces: we show that hydrophobic proteins typically have low expression, suggesting cells avoid an overabundance of sticky proteins.
\end{abstract}
\end{titlepage}

\section{Introduction}

\begin{figure*}[h!]
  \centering
  \includegraphics[width=\linewidth]{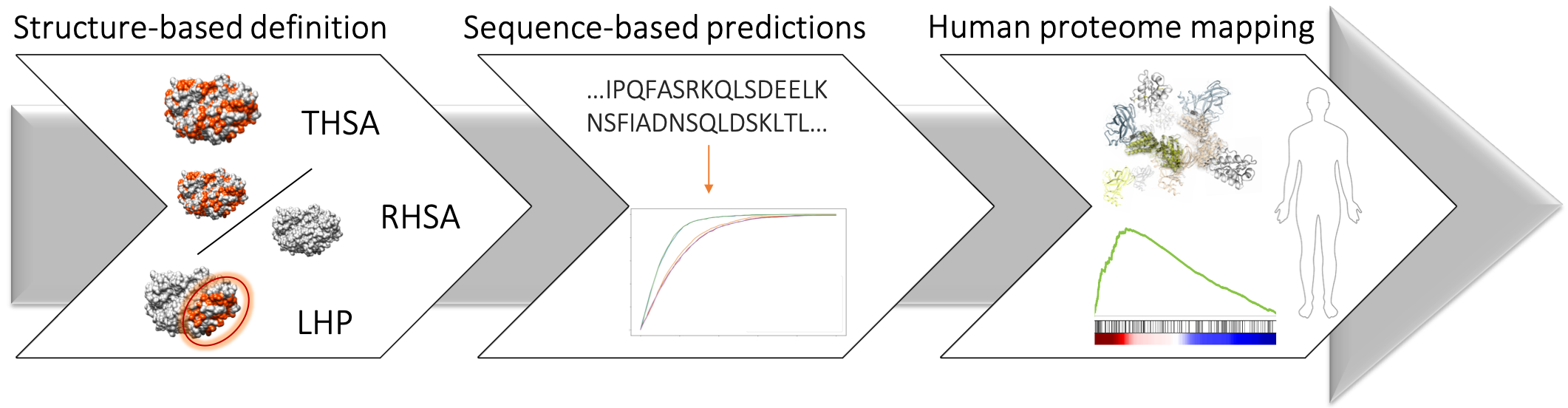}
  \caption{\textbf{Outline of the study}. 1) Structure-based definition represents the three hydrophobic measures: The red colour indicates the surface of hydrophobic residues. The total hydrophobic surface area (THSA) is calculated by summing the area of all hydrophobic residues. The relative hydrophobic surface area (RHSA) is calculated by dividing the THSA by the total accessible surface area (TASA). The largest hydrophobic patch surface area is the largest area of adjacent hydrophobic residues. 2) We train and benchmark sequence-based prediction methods of the three hydrophobic measures. 3) THSA, RHSA and LHP values for the human proteome were predicted by the best performing methods and used to estimate the abundance of hydrophobic proteins in various diseases and tissues.}
  \label{fig:intro_figure}
\end{figure*}

Hydrophobic residues tend to be buried inside the core of a protein to avoid contact with their hydrophilic surroundings (the hydrophobic effect)~\citep{Dill1985, Dill1990}. Hydrophobic residues that do occur on the protein surface often play a functional role, e.g. for protein-protein interactions and membrane binding~\citep{Malleshappa2014, Chothia1975,Young1994}. Additionally, exposed hydrophobic residues can play a role in the progression of diseases. For example, it has become apparent that hydrophobicity may play a major role in the formation and stabilisation of amyloid fibrils~\citep{Iadanza2018, Tuttle2016, Gils2020}, which are linked to aggregation diseases such as Alzheimer and Parkinson~\citep{Dobson2001,Koo1999, Ross2004,Chiti2006}. In fact, burying the hydrophobic residues inside the folded protein is also thought to prevent aggregation~\citep{Dobson2003, abeln2008disordered, abeln2011accounting}. Abundant exposed hydrophobic residues can also affect experimental outcomes: exposed hydrophobic residues may cause gel formation and prevent crystallisation for protein structure determination~\citep{Wright1999}; in  liquid chromatography surface hydrophobicity is used to separate proteins for further experiments~\citep{moruz2017peptide}. All these examples suggests that the more hydrophobic a protein surface, the more “sticky” this protein is to its surrounding (see also panel 1 in Figure~\ref{fig:intro_figure}). 

The hydrophobic surface area can be defined in different ways. Here, we use three different structure-based measures to describe surface hydrophobicity (see panel 1 in Figure~\ref{fig:intro_figure}):
\begin{enumerate}
  \item The total hydrophobic surface area (THSA) is the absolute area of all the exposed hydrophobic residues.
  \item The relative hydrophobic surface area (RHSA) is the fraction of the protein surface that is hydrophobic, i.e. the THSA divided by the total accessible surface area (TASA).
  \item The largest hydrophobic patch (LHP) is the largest connected hydrophobic area on the protein surface (and is therefore always smaller than or equal to the THSA). It has been shown that LHP size affects protein solubility~\citep{Lijnzaad1997, Bahadur2003, Huang2000} and function~\citep{Larsen1998, dobson2004principles}.
\end{enumerate}

Note that THSA, RHSA and LHP may not always correlate. For example, a large THSA value can be due to the size of the protein, and a protein with many scattered hydrophobic residues on its surface may have a small LHP but a large THSA and RHSA.

Experimentally, the exposed hydrophobic surface area can be estimated using differential scanning calorimetry (DSC), for which the heat capacity temperature relation for the folded protein is directly related to the THSA~\citep{Gomez1995,Dijk2016}.

In this work, our main goal is to investigate \emph{how} hydrophobic protein surfaces are within the human proteome. We also provide some insight how hydrophobicity is related to cellular expression levels, giving an idea of the overall hydrophobicity in the cellular environment. The question \emph{why} some of the human proteome is hydrophobic is not the main focus of our investigation, but is considered in some cases to interpret results. 

We use 3D structural information from the PDB to determine the THSA, RHSA and LHP from structure. The THSA and RHSA can be derived by summing over the exposed surface area per residue calculated by DSSP~\citep{kabsch1983dictionary}. To calculate the LHP, we introduce a novel method named MolPatch, which is loosely based on a method developed by Lijnzaad et al.~\citep{Lijnzaad1996, Lijnzaad1997}.

Since many protein structures have not yet been determined experimentally, we subsequently use the values we obtain from the PDB structures to train/assess predictors for these three measures. There is a wide range of methods that can predict the surface accessibility for a single residue~\citep{Garg2005, Petersen2009, Joo2012,Faraggi2012, Klausen2019}. However, to predict whether a \emph{hydrophobic} residue will be exposed to the surface is not a trivial task: the earlier methods tended to predict the majority of hydrophobic residues to be fully buried (see Figure~\ref{fig:benchmark_old}), as may be expected since the hydrophobicity of residues is strongly associated with being buried inside the protein~\citep{kyte1982}. The current generation of residue-based surface accessibility predictors use deep neural networks. For example, NetSurfP2.0 is a deep learning-based multitask predictor, which uses evolutionary profiles to make sequence-based predictions of structural features~\citep{Klausen2019}. It uses both convolutional and long short-term memory neural layers in the deep learning architecture, with the ability to predict both secondary structure and solvent accessibility~\citep{Klausen2019}. Here we will show NetSurfP2.0 is able to make accurate enough surface accessibility predictions for hydrophobic residues, which in turn can be used to predict the global hydrophobic surface measures described above. 

Finally, we use the best-performing prediction methods to predict the THSA, RHSA and LHP of all proteins in the human proteome, and correlate this to cellular expression levels, providing effectively an indication of proteome hydrophobicity per cell type. Subsequently, we use our predictions to provide a glance into the potential implications of a highly hydrophobic proteome in terms of human disease.

\section{Results}

\subsection{Structure-based definitions - MolPatch}
\begin{figure}[hbt!]
    \centering
    \includegraphics[width=\linewidth]{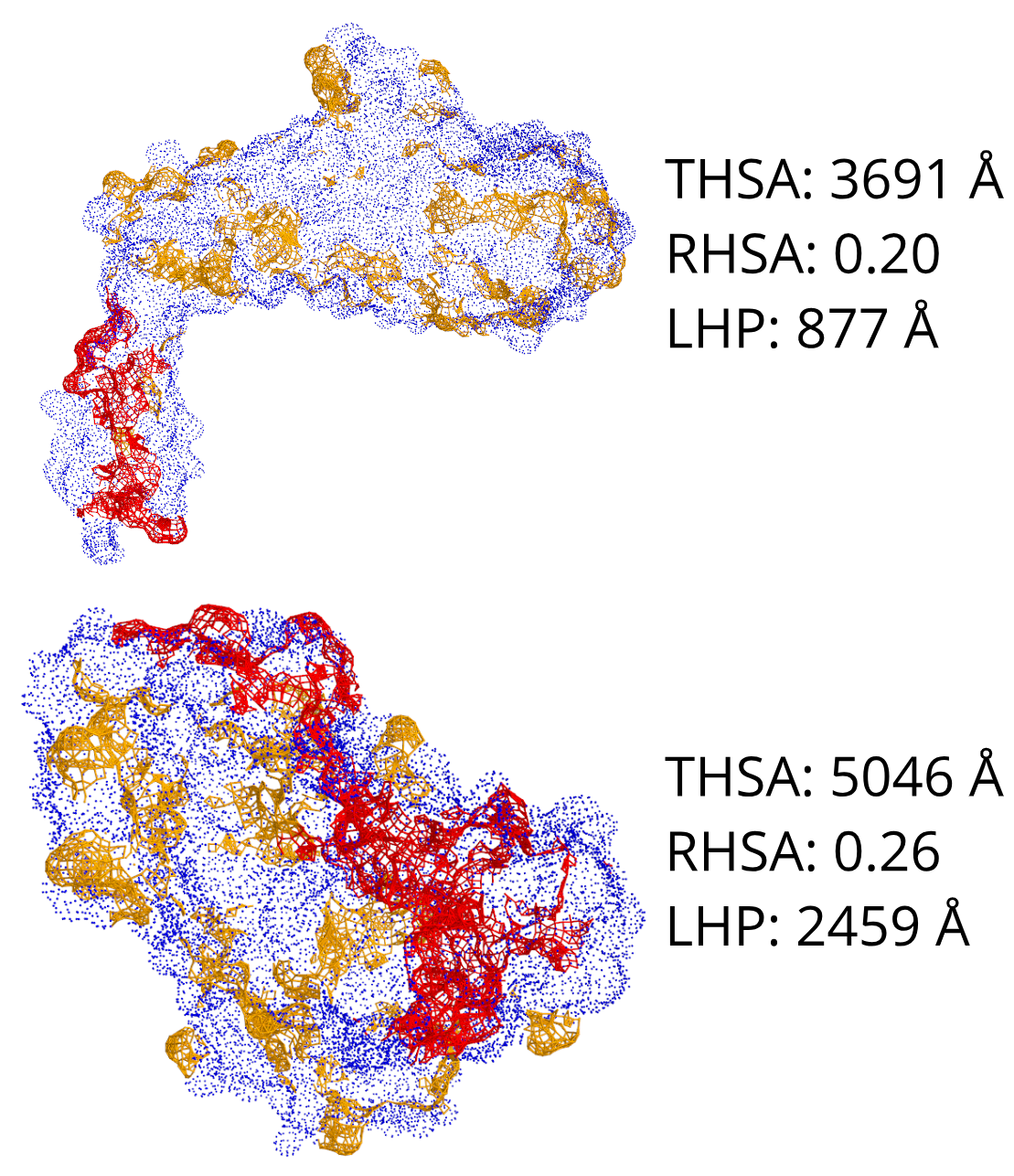}
    \caption{\textbf{THSA, RHSA and LHP, as identified by MolPatch for two different protein structures}. Top: SabA, PDB=4O5J. Bottom: Leishmanolysin, PDB=1LML. The surface of hydrophobic residues are displayed in yellow and red. Those in the largest hydrophobic patch (LHP) are displayed in red. The surfaces of the hydrophilic residues are displayed in blue. Note that Leishmanolysin is much larger (465 residues) and has a much larger THSA (5046 \AA) compared to SabA (370 residues, 3691 \AA), while the RHSA is quite similar between the two proteins, 26\% vs 20\%. The difference in the LHP is even larger, with 2459 \AA vs. 877 \AA, respectively; a nearly three-fold difference.}
    \label{fig:molpatch}
\end{figure}

To quantify the exposed hydrophobic areas on the protein surface, we defined three different structure-based measures for surface hydrophobicity, the THSA, RHSA and LHP. Using DSSP~\citep{kabsch1983dictionary}, we can calculate the THSA and RHSA directly from the surface area per residue (Figure~\ref{fig:molpatch}), see methods for futher details.

To define the largest hydrophobic patch on a protein surface, we developed a novel tool named MolPatch. This tool takes the 3D coordinates in PDB format and identifies networks of adjacent hydrophobic residues to find hydrophobic patches on the protein surface. Hydrophobic patches of 4,250 structures of soluble proteins were analysed using MolPatch (see Methods). Figure~\ref{fig:molpatch} highlights the importance of having three measures by observing the LHP of two proteins with very different surface areas. Although the difference in RHSA between the two proteins is only 6\%, the THSA and LHP of SabA are approximately 1.5 and 3 times larger than the LHP of Leishmanolysin. Generally, we see that there is no trivial correlation between THSA, RHSA and LHP (Figure~\ref{fig:Structure_based_scatterplot}).

To determine whether our structure-based largest patch definition is reasonable in biological terms, we overlapped the residues in the 20 largest hydrophobic patches of each protein in our database with those in the PiSITE protein interaction database (also see SI Methods). We would expect that large hydrophobic patches, functionally may serve as a protein-protein interaction interfaces. Indeed, we found that overall, the three largest patches in a protein were significantly enriched in protein interaction sites (Figure~\ref{fig:pisite} and Table~\ref{tab:pisite}).

\subsection{Sequence-based predictions - THSA and RHSA can be predicted with reasonable accuracy}
Since there are many more protein sequences available than structures, it is highly valuable to be able to predict the THSA, RHSA and LHP from sequence, which will allow us to characterise much broader set of proteins. Thus, we aimed to determine how well we can currently predict the three measures, and identify which sequence features contribute most to the accuracy of these predictions. We used our structure-based definition set to develop sequence-based predictors in a double cross-validation scheme (see Methods).

\begin{figure}[hbt!]
    \centering
    \includegraphics[width=.9\linewidth]{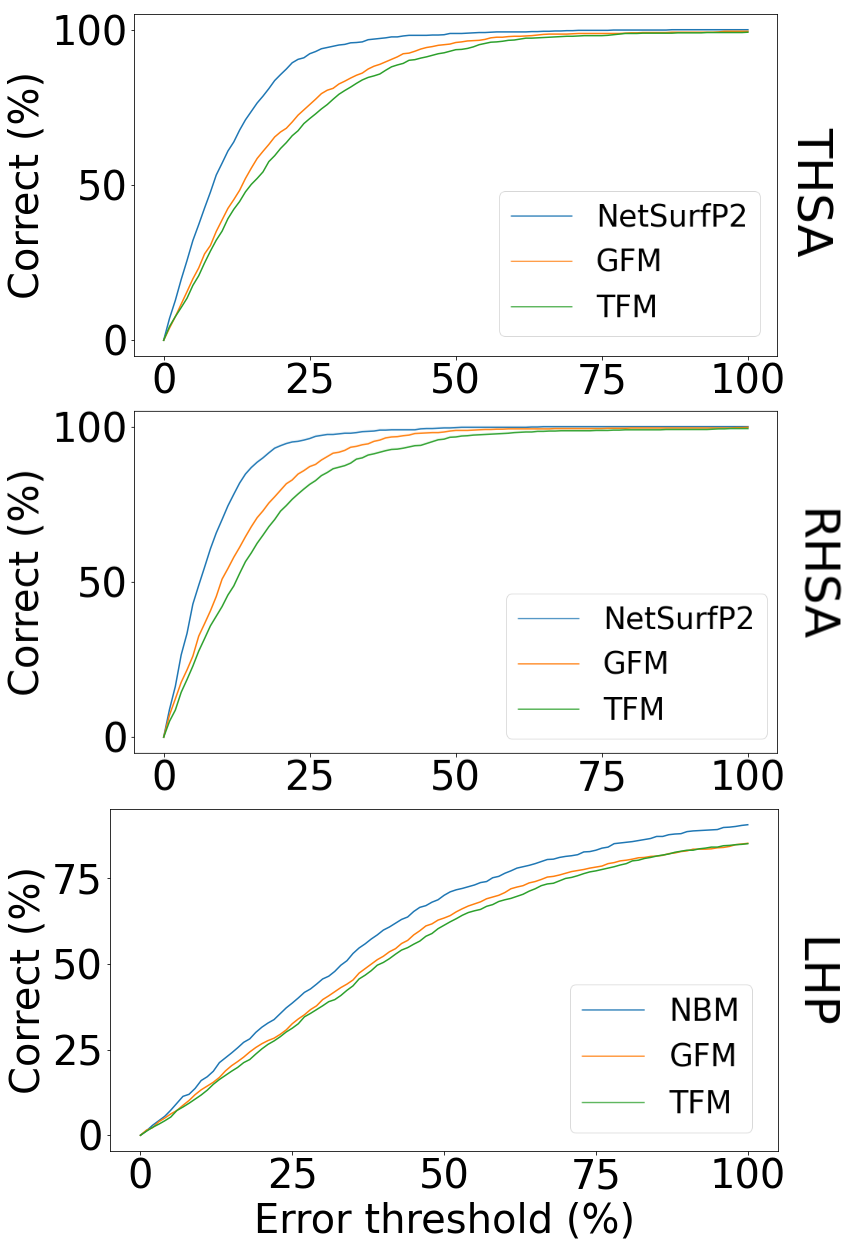}
    \caption{\textbf{Accuracy of the predictions of the total, relative and largest patch hydrophobic surface area for NetSurfP2.0-based models, the LBM, TFM and GFM.} The fraction of correctly predicted proteins within a certain error margin for each of the methods are shown as calculated over the test set.}\label{fig:accuracy}\vspace*{-10pt}
\end{figure}

To predict the THSA and RHSA, we used NetSurfP2.0, a neural-network-based method that takes evolutionary conservation profiles as input, and is currently one of the best (non-ensemble) predictors for surface accessibility and secondary structure~\citep{Klausen2019, xu2020opus, fereshteh2020enhancing}. NetSurfP2.0 provides surface area predictions per residue. To obtain the THSA, we summed over the predicted accessible surface areas of all hydrophobic residues. To obtain the RHSA, we summed over the predicted accessible surface area of all residues and divided the THSA by this value. Previous results (Figure~\ref{fig:benchmark_old}) indicate that the sequence length and hydrophobicity are strong predictors for the THSA and RHSA, and even outperformed a previous version of NetsurfP2.0 (Figure~\ref{fig:feature_importance}). Therefore, we trained two additional models, one that incorporates the sequence length, the number of hydrophobic residues and the number of hydrophilic residues (three-feature model, TFM), and one that includes a larger number of features derived from the sequence (global feature model (GFM) see Methods). Figure~\ref{fig:accuracy} and Table~\ref{tab:Rsquared} show that the NetSurfP2.0 based predictions are clearly superior.

The TFM, which only includes the features sequence length, number of hydrophobic and number of hydrophilic residues, also performs significantly better than random for both the THSA and RHSA, indicating that these features are of major significance for predicting these two properties. The GFM, which includes 31 features, performs only marginally better than the TFM, indicating that sequence length and sequence hydrophobicity are some of the main determinants for the hydrophobic surface area. Since it is difficult to obtain feature importance from neural network models such as NetSurfP2.0, we also analysed the feature importance measures from the GFM. This analysis showed that the hydrophobicity of the sequence is another major predictor for the THSA and RHSA (Figure~\ref{fig:feature_importance}, gravy score~\citep{kyte1982simple}, aromaticity~\citep{lobry1994hydrophobicity}, hydr\_count). 

To predict the LHP from sequence, the LHP determined by MolPatch was used as a gold standard. The training procedure for the TFM and GFM for predicting the LHP, was performed in a similar fashion to  the training for THSA and RHSA. Since the NetSurfP2.0 predictions cannot readily be used to predict the LHP, a model was trained that uses the THSA and RHSA predicted by NetSurfP2.0 as input features to predict the LHP (NetSurfP-based model, NBM). The results are shown in Figure~\ref{fig:accuracy} and Table~\ref{tab:Rsquared}. One can see that the NBM outperforms the other two methods. The sequence hydrophobicity again appears to have a major contribution to the prediction results (Figure ~\ref{fig:feature_importance}). Nevertheless, each of these prediction models perform significantly worse than the models for the THSA and RHSA predictions (Table~\ref{tab:Rsquared}), suggesting LHP prediction is less straightforward than THSA or RHSA predictions.

\begin{table*}[hbt!]
\caption{\label{tab:Rsquared} $R^2$ of each of the prediction models for the THSA, RHSA and LHP for the four different prediction models as calculated over the test set.}
\centering
{\begin{tabular}{@{}lllll@{}}\toprule & THSA & RHSA & LHP & Features \\\midrule
    NetsurfP2.0 & 0.92 & 0.77 & - & Evolutionary profiles \\
    NBM & - & - & 0.43 & THSA and RHSA predictions \\
    &&&& by NetSurfP2.0 \\
    TFM & 0.71 & 0.13 & 0.00 & Sequence length, number of  \\
    &&&& hydrophobic residues, number of \\
    &&&& hydrophilic residues \\
    GFM & 0.75 & 0.49 & 0.12 & 31 sequence-based features \\
\bottomrule
\end{tabular}}
\end{table*}

\subsection{Human Proteome Mapping}
\subsubsection{Transmembrane proteins - the most hydrophobic part of the human proteome}
\begin{figure*}[hbt!]
  \centering
  \includegraphics[width=\linewidth]{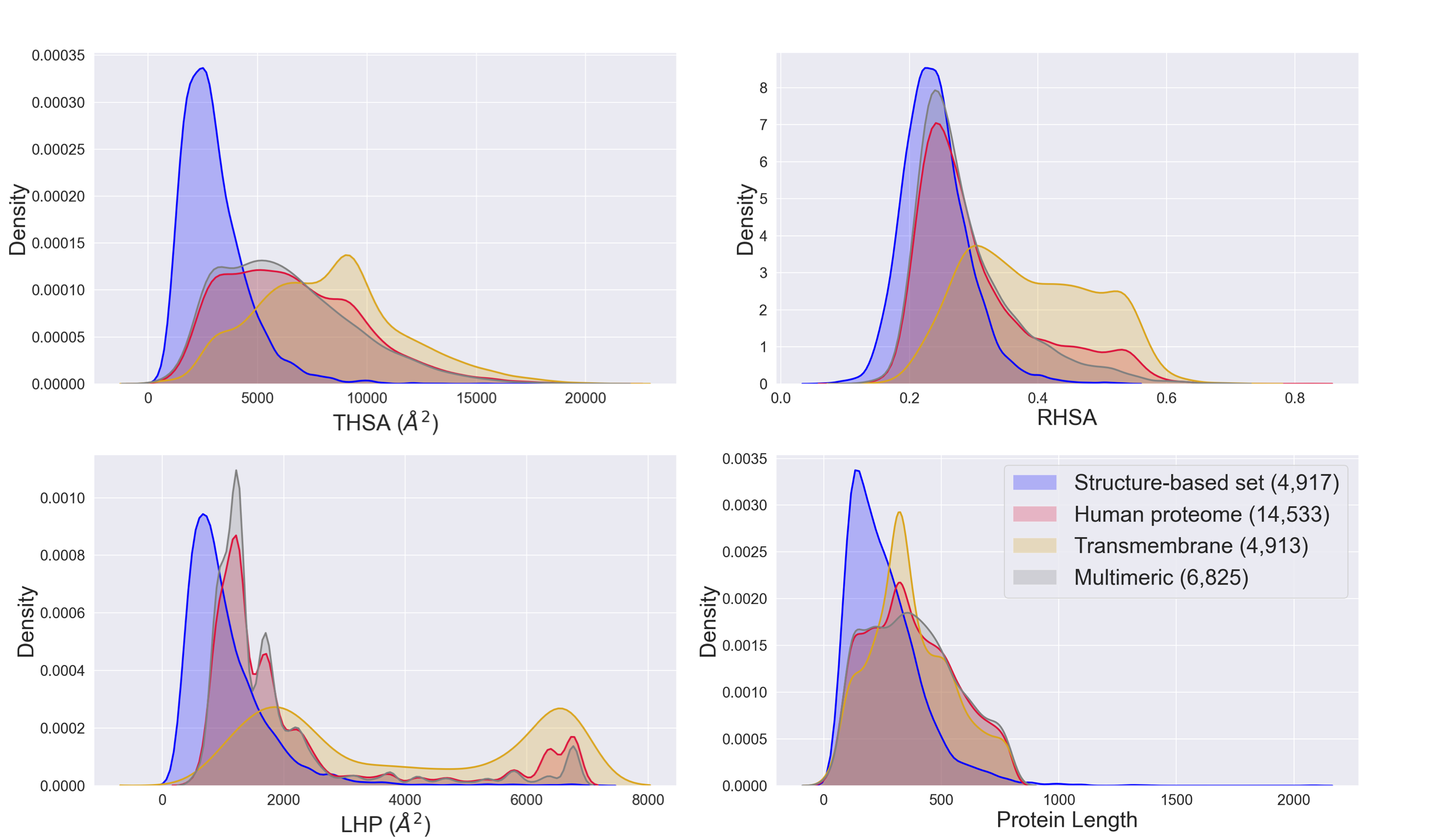}
  \caption{\textbf{The distribution of the protein length,  THSA, RHSA and LHP values from the whole curated human proteome (red), annotated transmembrane (yellow) and multimeric (grey) proteins (predicted) and the same values in the structure-based data set (blue) for the comparison.} The structure-based data set contains smaller proteins that the human proteome datasets, as may be expected. Values on the legend indicates the size of the data sets analysed. The figure indicates that transmembrane proteins are predicted to have large hydrophobic surface areas (observed on the LHP plot: $\sim 2000$ \AA; $\sim 6500$ \AA), which can be seen as a known positive control for the human proteome predictions.}
  \label{fig:Distribution}
\end{figure*}

For 14,533 proteins in the human proteome, we were able to predict THSA, RHSA and LHP values (see Methods). Figure~\ref{fig:Distribution} shows a comparison of the distributions of the definitions of these values on the structural data set and of the predicted values on the human proteome data set. Proteins in the structure-based data set appear to be smaller compared to those in the curated human proteome. In line with this, we see that the THSA and LHP distributions are strongly shifted towards the right-hand side compared to the structure-based data, most likely due to the larger size of proteins in the human proteome.

Moreover, the structure-based set (blue) does not show a peak of very large hydrophobic patches (LHP, $\sim 6500$ \AA) as observed for the human proteome data set (red). Importantly, structure-based data analysed by MolPatch neither contains proteins with more than one chain in the PDB structure nor transmembrane proteins; both groups of proteins maybe expected to have a very large hydrophobic patch. To investigate if this peak for the human proteome may be due to transmembrane or multimeric proteins, we selected those proteins annotated by UniProt~\citep{uniprot2019uniprot} as 'transmembrane' (yellow), or 'part of the protein complex' (grey). Indeed, after selecting transmebrane proteins from the human proteome data set, the composition of peak of the large hydrophobic patches, as well as the shoulder in the RHSA distribution can be explained predominantly through the transmembrane annotated proteins in the humen proteome (Figure~\ref{fig:Distribution}). Multimeric proteins mostly follow the distribution of the whole human proteome and do not appear to be much more hydrophobic in general. The results in Figure~\ref{fig:Distribution} also suggest that our ML model (NBM) successfully predicted transmembrane proteins to have large hydrophobic patches, despite the lack of transmembrane proteins in the training data set.

\subsubsection{Cells avoid the over expression of proteins with a large hydrophobic surface area}
\begin{figure*}[hbt!]
  \centering
  \includegraphics[width=\linewidth]{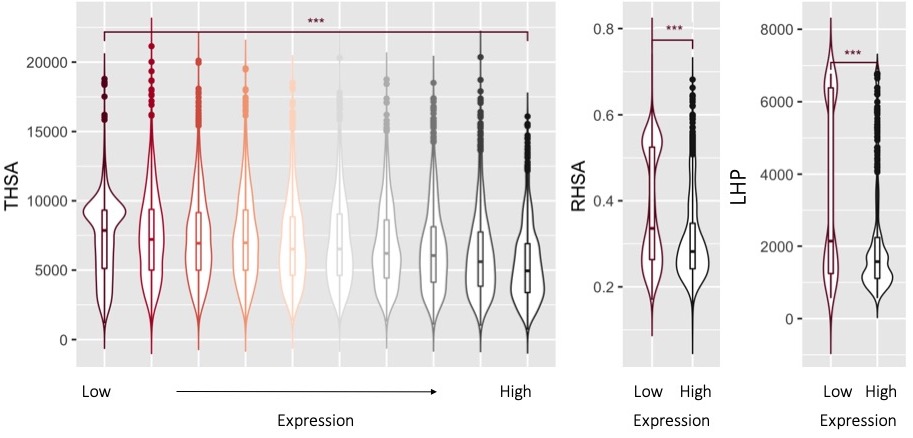}
  \caption{\textbf{Relationship between normalised expression (NX) and THSA, RHSA and LHP values.} For each gene the highest NX value was selected across all tissues.  The genes were grouped in deciles based on their expression levels. The groups with the lowest NX values were associated with signficantly higher THSA, RHSA and LHP values, compared to groups with the highest NX values. Significance was calculated using Wilcoxon signed-rank test.The three asteriks indicate p-values: $<$ 2.22e-16.}
  \label{fig:expression}
\end{figure*}

Since hydrophobic characteristics are associated with the aggregation tendencies, we wanted to investigate whether proteins with large hydrophobic surface areas have different expression levels. We used the RNA consensus tissue gene data from the Human Proteome Atlas to explore a link with expression levels. For this we relate normalised expression (NX) data to measures for surface hydrophobicity. To obtain a single expression value for each gene we took the highest expression level in any tissue. Figure~\ref{fig:expression} shows that the higher the expression level of the protein, the lower the THSA, RHSA and LHP value. 

We also explored the highly expressed genes based on a median NX value (across all the tissues that a particular gene appears in): these values show a similar trend (Figure~\ref{fig:expression_median}). Interestingly, proteins that do not follow the general trend, i.e. those that are highly expressed while having a large THSA, RHSA and LHP value, are typically protein subunits assembling large multimeric complexes. In such complexes the proteins are likely to be stably bound, and are hence able to shield the hydrophobic surfaces from the solvent.

\subsubsection{The brain- and kidney-specific proteomes are enriched with hydrophobic proteins}
To investigate if genes that are enriched in specific tissues are associated with the hydrophobic properties of the proteins, we carried out Gene Set Enrichment Analysis (GSEA). We downloaded 5 tissue enriched gene sets from the Human Protein Atlas (HPA)~\citep{ponten2008human, uhlen2015tissue}. Table~\ref{tab:GSEA_HPA} shows that the brain and kidney tissue-enriched gene sets have a high enrichment in predicted THSA and LHP values. 
Kidney-enriched genes show the highest enrichment in THSA, RHSA and LHP of the ranked gene lists (p-values < 0.001). A possible explanation for this is the major role of kidney tissue in maintaining homeostasis through various membrane-bound receptors and transporters~\citep{ lote1994principles}. Indeed, 79\% of the kidney enriched proteome is annotated as transmembrane by UniProt~\citep{uniprot2019uniprot}. Interestingly, liver tissue revealed no enrichment. The skin and blood tissue enriched gene sets exhibited significant enrichment in the RHSA ranked list. Furthermore, both tissue groups were significantly depleted in the THSA ranked list, indicating that they may contain the smaller proteins in the human proteome.

\begin{table*}[h!]
\caption{\label{tab:GSEA_HPA} \textbf{Pre-ranked GSEA enrichment statistics in different tissues.} Various tissue-enriched gene sets were obtained from the HPA~\citep{ponten2008human, uhlen2015tissue}. THSA, RHSA and LHP values were central-scaled prior to the GSEA analysis. The enrichment score (ES) is the maximum deviation from zero showing the degree to which the gene set is over-represented at the top (positive ES score) or bottom (negative ES score) of the entire ranked list of genes. The fraction of transmembrane and multimeric proteins in the following gene sets is shown in percentages.
* P < 0.05 
**  P < 0.001}
{\begin{tabular}{@{}llllll@{}}\toprule Gene set & ES (THSA) & ES (RHSA) & ES (LHP) & Transmembrane (\%) & Multimeric (\%) \\\midrule
    Brain (488) & 0.33$^{**}$ & 0.14 & 0.64$^{**}$ & 47.0 & 47.0 \\
    Kidney (53) & 0.62$^{**}$ &  0.53$^{**}$ & 0.78$^{**}$ & 79.2 & 35.8 \\
    Skin (113) & -0.46$^{*}$ & 0.30$^{**}$ & 0.44 & 7.9 & 15.9 \\
    Liver (242) & -0.22 & -0.16 & 0.45 & 26.0 & 59.9 \\
    Blood (57) & -0.41$^{*}$ & 0.40$^{**}$ & 0.68$^{*}$& 47.4 &  28.0 \\
\bottomrule
\end{tabular}}
\end{table*}

To investigate the overall tissue hydrophobicity, we introduced TASH - tissue specific average surface hydrophobicity for \emph{all proteins} based on the expression levels in a specific tissue (Eqn. \ref{eq:TASH} and Figure~\ref{fig:TASH}). TASH-THSA value provides an indication of the total hydrophobic surface area present in a specific cell type. The tissues with the highest TASH-THSA values occur in the brain, such as the cerebellum, corpus callosum, thalamus, cerebral cortex, and basal ganglia (Figure~\ref{fig:TASH}). 

\subsubsection{Increased relative hydrophobicity is associated with (aggregation) diseases}
To investigate the association of surface hydrophobicity with human diseases, a GSEA pre-ranked analysis of 375 various disease-associated gene sets was carried out, of which 44 gene sets show a significant (p-value < 0.05) enrichment (< -0.2 (negative enrichment) and >0.2 (positive enrichment) in at least two hydrophobic measures (see Figure~\ref{fig:GSEA_Diseases}). Among the enriched gene sets we can observe several KEGG~\citep{kanehisa2000kegg} pathways that are associated with neurological disorders. The RHSA showed a significant (p-value < 0.05) enrichment in Parkinson's (ES=0.43), Alzheimer's (ES=0.24) and Huntington’s disease (ES=0.23) gene sets. The analysis shows a significant (p-value < 0.001) enrichment of sticky proteins (based on LHP) in the KEGG Parkinson's disease map (ES=0.66). In contrast to the GSEA analysis results on tissue-specific proteome, the THSA shows a negative enrichment in these sets, suggesting that the proteins involved in pathological pathways have large hydrophobic surfaces and patches, but are smaller in size (median length 171-180 residues).

\section{Discussion}
In this work, we analysed the predictability of hydrophobic areas on protein surfaces, which until recently was a difficult problem. We show that THSA and RHSA values can be predicted with high accuracy (>75\% within a 20\% error margin, Figure~\ref{fig:accuracy}). The improved predictions of NetSurfP2.0, compared to the earlier secondary structure prediction methods (Figure~\ref{fig:benchmark_old}), make this possible by straightforward calculations of the THSA and RHSA using the predictions of the surface accessibility per residue from NetSurfP2.0. On the other hand, the LHP cannot be directly obtained from NetSurfP2.0~\citep{Klausen2019} and needs additional model training. Nevertheless, we believe that recent advances in deep neural nets, contact map prediction and structure prediction~\citep{zheng2019deep, li2019respre, xu2020opus, senior2020improved} should make it possible to make these predictions more accurate in the near future, for example by using structure or contact predictions to predict the hydrophobic patches, or by training a purpose specific deep neural net. 

When investigating the link between tissue-based expression levels and the measures for surface hydrophobicity, we clearly observe that highly expressed proteins typically do not have a large hydrophobic surface area (THSA, RHSA and LHP as seen in Figure~\ref{fig:expression}). A similar trend has previously been observed for proteins with a strong tendency to form amyloid fibrils~\citep{tartaglia2009}, suggesting an evolutionary pressure to avoid proteins with high aggregation propensities being present at high concentrations in the cell. Based on our data, if we assume that the high expression values correlate with high protein abundance in the cell, it is conceivable that there is also an evolutionary pressure against proteins with a large hydrophobic surface area to be overly abundant in the cell. 

Note that while the THSA and RHSA sequence-based predictions show a reasonable correlation with the structure-based definitions, this does not necessarily mean that the predicted amount of hydrophobic surface accessible area is actually exposed to the cellular environment. For example, a hydrophobic patch may be buried in a stable macro-molecular complex, or may be buried inside a membrane. Additionally, a high hydrophobic surface area does not necessarily mean a protein will be insoluble; this will also be very much dependent on the amount of polar and charged residues that may surround the hydrophobic residues or patches~\citep{kramer2012toward}, as well as disordered regions~\citep{abeln2008disordered}.

Despite the general tendency to avoid highly expressed proteins with a large hydrophobic surface area, the brain appears to be highly hydrophobic in its overall expression patterns (THSA in cerebellum, cerebral and cortex as shown in Figure~\ref{fig:TASH}) and in proteins enriched in the brain (THSA and LHP as shown in Table~\ref{tab:GSEA_HPA}). This high expression of proteins with a large hydrophobic surface area may be rationalised by functional requirements: genes enriched in brain tissue are involved in organising and maintaining synaptic signalling, requiring various cell adhesion proteins with large hydrophobic surface areas~\citep{sytnyk2017neural}; the cellular morphology of neurons including the dendrite means that there is a relatively large transmembrane surface area per cell. Additionally, the structural integrity of neuronal axons is facilitated by myelin~\citep{stadelmann2019myelin}, a fatty substance surrounding neurons, and by myelin-associated proteins, which are all very hydrophobic.

Furthermore, brain tissue has been associated with various aggregation diseases~\citep{Dobson2001, Koo1999, Ross2004, Chiti2006}. Based on our data, it may be hypothesised that the brain is specifically vulnerable to such diseases due to its high expression of proteins with a large hydrophobic surface. Hydrophobic patches play a role in the folding and/or misfolding of proteins~\citep{dobson2004principles, Ross2004}, and can possibly provide nucleation sites for the formation of oligomers and amyloid fibrils. This hypothesis would be supported by the relatively high hydrophobic surface area in molecular pathways associated with Parkinson's, Huntington's and Alzheimer's disease (as observed for the RHSA and LHP, see Figure~\ref{fig:GSEA_Diseases}).

\section{Conclusion}
In summary, we defined measures for surface hydrophobicity: THSA, RHSA and LHP. For the definition of the LHP, we created a new tool, MolPatch, that can identify the LHP of a protein given its PDB structure. Secondly, we have shown that the THSA and RHSA can be predicted with high accuracy by adapting the output of NetSurfP2.0, whereas the LHP is more difficult to predict using currently existing methods. Finally, we showed that a high hydrophobicity of a protein surface is associated with lower general expression levels, suggesting that evolutionary pressure keeps the abundance of such proteins low. In addition, we show that brain tissue expresses relatively many proteins with a large hydrophobic surface area, giving a possible explanation for why the brain is relatively prone to diseases that are associated with misfolding and aggregation. 

\section{Methods and Materials}
\begin{figure*}[tbh!]
  \centerline{\includegraphics[width=0.95\textwidth]{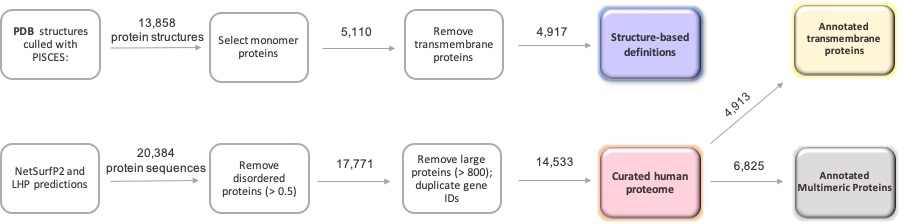}}
  \caption{\textbf{Data curation scheme representing the main steps used to generate data sets for this study.} The boxes show the filtering steps and the arrows indicate the number of entries (structures or sequences) passed through. The structure-based definitions data set used the protein 3D structure information and the human proteome data set was constructed of protein sequences. The distribution of the measures for surface hydrophobicity within the data sets are represented by the Figure~\ref{fig:Distribution} and are colour-coded.}\label{fig:data_curation}
\end{figure*}

Figure~\ref{fig:intro_figure} indicates how our approach is split into three stages. Firstly, we created a database of filtered PDB structures (Figure~\ref{fig:data_curation}) using PISCES. We used this culled set to define measures for surface hydrophobicity: THSA, RHSA and LHP. For the latter, we used a newly developed tool named MolPatch. Secondly, using the same dataset, we investigated how well we can predict these measures from sequence using the output generated by NetSurfP2.0. Finally, we determined the biological impact of the THSA, RHSA and LHP. To this end, we created a dataset of human proteins in Uniprot~\citep{uniprot2019uniprot}. We used the best prediction models to predict the THSA, RHSA and LHP for each of these proteins. Subsequently, we correlated gene expression to the hydrophobicity in the human proteome for different cell types. 

\subsection{Introducing measures for hydrophobicity}
We define the Total Hydrophobic Surface Area (THSA) as the sum of the surface areas of all hydrophobic residues in the protein. For proteins with an available 3D structure in the PDB, this quantity can be determined by calculating the surface area of each residue using DSSP (we used the DSSP module in Biopython version 1.76)~\citep{Cock2009}. The Relative Hydrophobic Surface Area (RHSA) is the caclulated as the THSA was divided by the total surface area of all residues in the protein. Residues, $r$, were considered hydrophobic in this work, if: $r \in \left\{ A,C,F,I,L,M,V,W,Y \right\}$

In order to calculate the surface area of the Largest Hydrophobic Patch (LHP) based on a protein structure, we need to find the largest connected hydrophobic surface area on a protein surface. For this purpose, we developed the tool MolPatch. Given the PDB structure of a protein, MolPatch creates a point cloud on the solvent-excluded protein surface (SES) using MSMS~\citep{sanner1996reduced}. In this work, the SES was constructed using a probe of 1.5 \AA~ and a density of 1.5 points per \AA$^2$. Each point on the point surface was labelled hydrophobic or hydrophilic based on the hydrophobicity classification of the closest \emph{residue}. Initial edges between points were then created if the points existed within a range of 1.25\AA of each other. This search was performed with the KDTree algorithm to speed up the process~\citep{Bentley1975}. Finally, only the edges between hydrophobic labeled node pairs were retained. This created a network of isolated hydrophobic patches. The individual network components were then extracted for accessible surface area estimation. MolPatch can also carry out hydrophobic patch identification using atom-based definitions of hydrophobicity for each SES point rather than residue-based definitions, as available on GitHub. In this work we only use the residue-based method.

\subsection{Sequence-based predictions}
\subsubsection{Data curation}
To predict the THSA, RHSA and LHP, a dataset of PDB structures was generated using PISCES. PISCES is a public server for culling sets of protein sequences from the Protein Data Bank (PDB) by sequence identity and structural quality criteria~\citep{wang2003pisces}. This is important, because using structures with a high sequence identity can introduce bias in the dataset, and factors such as the resolution can affect the accuracy of the results. The chosen parameters were as follows: sequence percentage identity lower or equal to 25\%, resolution lower or equal to \SI{3.0}{\angstrom}, R-factor lower or equal to 0.3, sequence length within the range of 40-10,000 amino acids,  and non-X-ray entries and C$\alpha$-only entries were excluded. The culled data set consisted of 13,858 unique protein structures with a selection of 14,604 chains. Two obsolete PDB chains were removed. Multimeric proteins were filtered out, which resulted in a data set of 5,110 unique monomeric protein structures. Transmembrane proteins have a relatively large hydrophobic surface area. This clashes with our model which was made for soluble proteins where hydrophobic residues tend to be buried. TMHMM~\citep{sonnhammer1998hidden, moller2001evaluation} was used to filter transmembrane proteins from the data set (>18 amino acids in transmembrane helices).

\subsubsection{Machine Learning models}
The final data set for training and testing of the models contained 4,917 monomers. For the THSA and RHSA, the values calculated by DSSP (as described above) were used as training output labels. MolPatch was used to create training output labels for evaluating the LHP predictions. Predictions for the THSA, RHSA and LHP were acquired with the following models:

\begin{enumerate}
    \item The \textbf{three feature model (TFM)} uses the sequence length, number of hydrophobic amino acids and number of hydrophilic amino acids as input. This model is trained using a cubist regression in the CARET module~\citep{Kuhn2008}.
    \item The \textbf{global feature model (GFM)} uses 31 global features (counts of each of the 20 amino acids, sequence length, entropy, hydrophobic amino acid count, polar amino acid count, molecular weight, aromaticity, instability index, gravy score, buried, isoelectric point and molar extinction coefficient) as input. This model is trained using an XGBoost regressor~\citep{chen2016}.
    \item (THSA and RSHA only) \textbf{NetSurfP2.0} was used to predict the accessible surface area of all the amino acids in a protein. Subsequently, the THSA was calculated by summing over the predicted surface areas of the hydrophobic residues in the protein sequence. The RHSA was calculated by dividing the predicted THSA by the sum of the surface areas of all the residues in the protein as predicted by NetSurfP2.0.
    \item (LHP only) Since NetSurfP2.0 cannot predict the LHP directly, an XGBoost regressor model was trained using the RHSA and THSA predicted by NetSurfP2.0 (as described above) as input features. This model was called the \textbf{NetSurfP-based model (NBM)}.
\end{enumerate}

To assess the models, a double cross-validation loop was used. The data was randomly split into a training and test set of 80\% and 20\%, respectively. The training set was used to deploy a three-fold cross-validation scheme, in which the parameters for each of the models were optimised using a grid search method (code available on GitHub).
The final accuracy estimates were calculated over the test set.

\subsubsection{Estimation of prediction errors}
In order to evaluate the predictions, the structure-based definitions and sequence-based predictions can be compared, by calculating the correlation coefficient $R^2$. Nevertheless, for difficult regression tasks this value will put a lot of weight on the outliers, and will not produce results that are easy to interpret. In addition to the $R^2$ measure, we also evaluated the performance of the prediction model by examining the relative error threshold curve given a certain threshold, partially inspired by the GDT\_TS score~\citep{Zemla2001}. A major benefit of this method is that it is robust against extreme outliers. 
For each prediction, the relative THSA error ($\delta_{THSA_i}$), RHSA error ($\delta_{RHSA_i}$), and
LHP error ($\delta_{LHP_i}$) for each protein $i$ are defined by the following formulas:
\begin{equation}
    \delta_{THSA_i} = \frac{\left| THSA_{pred_i} - THSA_{DSSP_i} \right|}{THSA_{DSSP_i}}
    \label{eq:THSA_error}
\end{equation}
\begin{equation}
    \delta_{RHSA_i} = \frac{\left| RHSA_{pred_i} - RHSA_{DSSP_i} \right|}{RHSA_{DSSP_i}}
    \label{eq:RHSA_error}
\end{equation}
\begin{equation}
    \delta_{LHP_i} = \frac{\left| LHP_{pred_i} - LHP_{MolPatch_i} \right|}{LHP_{MolPatch_i}}
    \label{eq:LHP_error}
\end{equation}
\noindent where $THSA_{pred_i}$, $RHSA_{pred_i}$, and $LHP_{pred_i}$ are the predicted THSA, RHSA, and LHP of a protein. $THSA_{DSSP_i}$ and $RHSA_{DSSP_i}$ are the THSA and RHSA of a protein estimated using DSSP. $LHP_{MolPatch_i}$ is the predicted LHP of a protein, determined by MolPatch. The performance of the methods over the whole set of structures is evaluated by plotting the percentage correctly predicted instances (protein chains) versus a varying error threshold $t$. The threshold curve, $F(t)$, shows the percentage of correctly predicted THSA and RHSA of proteins for a given relative error threshold,$t$:
\begin{equation}
    F \left( t \right) = \frac{\left| \{ i | i \in chains \wedge \delta < t \} \right|}{\left| \{ i | i \in chains \} \right|} \cdot 100
    \label{eq:fraction_correct}
\end{equation}
The relative error for all chains in the chain dataset is calculated to determine the fraction of correctly predicted chains for the threshold, see also Figure~\ref{fig:error_example}. The $\delta$ is here interchangeably used for $\delta_{THSA_i}$, $\delta_{RHSA_i}$, or $\delta_{LHP_i}$. Unlike in a ROC-curve, the amount of correctly predicted chains does not necessarily have to be 100\% when the threshold $t=1.0$, since the size of the relative error can be $>100\%$. 

\subsection{Human proteome mapping}
\subsubsection{Data curation}
All reviewed protein sequences for the human genome were extracted from UniProt~\citep{uniprot2019uniprot} (accessed 1st Oct 2020). In total 20,384 sequences were analysed with NetSurfP2.0 for predicting solvent accessibility and structural disorder among other characteristics. THSA and RHSA values were calculated from NetSurfP2.0 predictions as described above. The LHP for each protein has been predicted using the NBM. The following data curation steps have been administered to remove unreliable predictions: (1) highly disordered proteins have been discarded (more than a half of the residues have been classified as disordered); (2) large proteins (> 800 AA residues) have been discarded in order to match the protein sizes in the structure-based definitions data set. (3) duplicate gene IDs were filtered out and the ones with the highest THSA value were retained. This quality filter resulted in a curated data set of 14,533 proteins. Seperate data sets were created with 4,913 proteins annotated as transmembrane and 6,825 - as multimeric by UniProt (Figure~\ref{fig:data_curation}).

Additionally, the final curated data set described above was used to analyse the link between the expression levels and measures for surface hydrophobicity. RNA consensus tissue gene data was downloaded from Human Protein Atlas~\citep{ponten2008human, uhlen2015tissue} (accessed on \url{https://www.proteinatlas.org/about/download} 24 Dec 2020). 

\subsubsection{Gene set enrichment analysis}
THSA, RHSA, and LHP values were centered (such that 0 fell between two parts of a bimodal distribution or between the main bulk and the tail of the distribution, see Figure~\ref{fig:centering}) and scaled (~\ref{eq:centering}) prior to the pre-ranked GSEA analysis ~\citep{subramanian2005gene, mootha2003pgc}. Tissue-enriched gene sets were downloaded from the Human Protein Atlas (accessed 10 Nov 2020). 375 Disease associated gene sets were extracted from the GSEA website (accessed on \url{https://www.gsea-msigdb.org/gsea/msigdb/search.jsp} 5 Nov 2020). GSEA was used with the following parameters: number of permutations = 1000; collapse; chip platform: human UniProt IDs MSigDB.v7.2.chip"; enrichment statistic: weighted; max size=1000, min size=15.

\subsubsection{Tissue-specific average surface hydrophobicity}
Tissue-specific average surface hydrophobicity (TASH) was calculated across all the genes with the following formula with and without transmembrane proteins: 

\begin{equation}
  \text{TASH}_t = \frac{ \sum_g  \text{NX}_{g,t} \cdot h_g}{\sum_g \text{NX}_{g,t}}
  \label{eq:TASH}
\end{equation}

\noindent Where $\text{TASH}_t$ is the tissue-specific average surface hydrophobicity for tissue $t$, $\text{NX}_{g,t}$ is the normalised expression of gene $g$ in tissue $t$ and $h$ is the predicted hydrophobicity of gene $g$ for one of the three measures (THSA, RHSA or LHP). The results are shown in Figure \ref{fig:TASH}.

\section{Author contributions}
JvE, RB and EvD performed the benchmark analyses for the prediction methods. JvE developed MolPatch. DG and JvG performed the expression and enrichment analyses. JvG, DG and SA wrote the manuscript. SA supervised and oversaw the project and was responsible for conceptualisation and funding acquisition. All authors reviewed the manuscript.

\section{Acknowledgments}
We would like to thank Dr. Bent Petersen for providing us with NetSurfP2.0 predictions for human genome data set, Prof. Jaap Heringa and Dr. Bernd Brandt for helpful discussion.

\section{Funding}
JvG and SA thank the Nederlandse Organisatie voor Wetenschappelijk Onderzoek (\url{https://www.nwo.nl/over-nwo/organisatie/nwo-onderdelen/enw})
for funding under project number 680-91-112 (NWO). DG and SA received funding from the European Union’s Horizon 2020 research and innovation programme under the Marie Skłodowska-Curie grant agreement No 860197, the MIRIADE project. RB has received funding from the Vlaams Agentschap Innoveren en Ondernemen under project number HBC.2020.2205.

\bibliography{refs_arxiv}

\begin{thebibliography}{10}
\expandafter\ifx\csname url\endcsname\relax
  \def\url#1{\texttt{#1}}\fi
\expandafter\ifx\csname urlprefix\endcsname\relax\def\urlprefix{URL }\fi
\providecommand{\bibinfo}[2]{#2}
\providecommand{\eprint}[2][]{\url{#2}}

\bibitem{Dill1985}
\bibinfo{author}{Dill, K.~A.}
\newblock \bibinfo{title}{Theory for the folding and stability of globular
  proteins}.
\newblock \emph{\bibinfo{journal}{Biochemistry}} \textbf{\bibinfo{volume}{24}},
  \bibinfo{pages}{1501--1509} (\bibinfo{year}{1985}).

\bibitem{Dill1990}
\bibinfo{author}{Dill, K.~A.}
\newblock \bibinfo{title}{Dominant forces in protein folding}.
\newblock \emph{\bibinfo{journal}{Biochemistry}} \textbf{\bibinfo{volume}{29}},
  \bibinfo{pages}{7133--7155} (\bibinfo{year}{1990}).

\bibitem{Malleshappa2014}
\bibinfo{author}{Gowder, S.~M.}, \bibinfo{author}{Chatterjee, J.},
  \bibinfo{author}{Chaudhuri, T.} \& \bibinfo{author}{Paul, K.}
\newblock \bibinfo{title}{Prediction and analysis of surface hydrophobic
  residues in tertiary structure of proteins}.
\newblock \emph{\bibinfo{journal}{The Scientific World Journal}}
  \textbf{\bibinfo{volume}{2014}} (\bibinfo{year}{2014}).

\bibitem{Chothia1975}
\bibinfo{author}{Chothia, C.} \& \bibinfo{author}{Janin, J.}
\newblock \bibinfo{title}{Principles of protein–protein recognition}.
\newblock \emph{\bibinfo{journal}{Nature}} \textbf{\bibinfo{volume}{256}},
  \bibinfo{pages}{705--708} (\bibinfo{year}{1975}).

\bibitem{Young1994}
\bibinfo{author}{Young, L.}, \bibinfo{author}{Jernigan, R.~L.} \&
  \bibinfo{author}{Covell, D.~G.}
\newblock \bibinfo{title}{A role for surface hydrophobicity in
  protein‐protein recognition}.
\newblock \emph{\bibinfo{journal}{Protein Science}}
  \textbf{\bibinfo{volume}{3}}, \bibinfo{pages}{717--729}
  (\bibinfo{year}{1994}).

\bibitem{Iadanza2018}
\bibinfo{author}{Iadanza, M.~G.} \emph{et~al.}
\newblock \bibinfo{title}{{The structure of a $\beta$2-microglobulin fibril
  suggests a molecular basis for its amyloid polymorphism}}.
\newblock \emph{\bibinfo{journal}{Nature Communications}}
  \textbf{\bibinfo{volume}{9}}, \bibinfo{pages}{4517} (\bibinfo{year}{2018}).
\newblock \urlprefix\url{http://www.nature.com/articles/s41467-018-06761-6}.

\bibitem{Tuttle2016}
\bibinfo{author}{Tuttle, M.~D.} \emph{et~al.}
\newblock \bibinfo{title}{{Solid-state NMR structure of a pathogenic fibril of
  full-length human $\alpha$-synuclein}}.
\newblock \emph{\bibinfo{journal}{Nature Structural {\&} Molecular Biology}}
  \textbf{\bibinfo{volume}{23}}, \bibinfo{pages}{409--415}
  (\bibinfo{year}{2016}).
\newblock \urlprefix\url{http://www.nature.com/articles/nsmb.3194}.

\bibitem{Gils2020}
\bibinfo{author}{van Gils, J. H.~M.} \emph{et~al.}
\newblock \bibinfo{title}{The hydrophobic effect characterises the
  thermodynamic signature of amyloid fibril growth}.
\newblock \emph{\bibinfo{journal}{PLOS Computational Biology}}
  \textbf{\bibinfo{volume}{16}}, \bibinfo{pages}{1--25} (\bibinfo{year}{2020}).
\newblock \urlprefix\url{https://doi.org/10.1371/journal.pcbi.1007767}.

\bibitem{Dobson2001}
\bibinfo{author}{Dobson, C.~M.}
\newblock \bibinfo{title}{The structural basis of protein folding and its links
  with human disease}.
\newblock \emph{\bibinfo{journal}{Philosophical Transactions of the Royal
  Society of London. Series B: Biological Sciences}}
  \textbf{\bibinfo{volume}{356}}, \bibinfo{pages}{133--145}
  (\bibinfo{year}{2001}).

\bibitem{Koo1999}
\bibinfo{author}{Koo, E.~H.}, \bibinfo{author}{Lansbury, P.~T.} \&
  \bibinfo{author}{Kelly, J.~W.}
\newblock \bibinfo{title}{Amyloid diseases: abnormal protein aggregation in
  neurodegeneration}.
\newblock \emph{\bibinfo{journal}{Proceedings of the National Academy of
  Sciences}} \textbf{\bibinfo{volume}{96}}, \bibinfo{pages}{9989--9990}
  (\bibinfo{year}{1999}).

\bibitem{Ross2004}
\bibinfo{author}{Ross, C.~A.} \& \bibinfo{author}{Poirier, M.~A.}
\newblock \bibinfo{title}{Protein aggregation and neurodegenerative disease}.
\newblock \emph{\bibinfo{journal}{Nature medicine}}
  \textbf{\bibinfo{volume}{10}}, \bibinfo{pages}{S10--S17}
  (\bibinfo{year}{2004}).

\bibitem{Chiti2006}
\bibinfo{author}{Chiti, F.} \& \bibinfo{author}{Dobson, C.~M.}
\newblock \bibinfo{title}{Protein misfolding, functional amyloid, and human
  disease}.
\newblock \emph{\bibinfo{journal}{Annu. Rev. Biochem.}}
  \textbf{\bibinfo{volume}{75}}, \bibinfo{pages}{333--366}
  (\bibinfo{year}{2006}).

\bibitem{Dobson2003}
\bibinfo{author}{Dobson, C.~M.}
\newblock \bibinfo{title}{Protein folding and disease: a view from the first
  horizon symposium}.
\newblock \emph{\bibinfo{journal}{Nature Reviews Drug Discovery}}
  \textbf{\bibinfo{volume}{2}}, \bibinfo{pages}{154--160}
  (\bibinfo{year}{2003}).

\bibitem{abeln2008disordered}
\bibinfo{author}{Abeln, S.} \& \bibinfo{author}{Frenkel, D.}
\newblock \bibinfo{title}{Disordered flanks prevent peptide aggregation}.
\newblock \emph{\bibinfo{journal}{PLoS Comput Biol}}
  \textbf{\bibinfo{volume}{4}}, \bibinfo{pages}{e1000241}
  (\bibinfo{year}{2008}).

\bibitem{abeln2011accounting}
\bibinfo{author}{Abeln, S.} \& \bibinfo{author}{Frenkel, D.}
\newblock \bibinfo{title}{Accounting for protein-solvent contacts facilitates
  design of nonaggregating lattice proteins}.
\newblock \emph{\bibinfo{journal}{Biophysical journal}}
  \textbf{\bibinfo{volume}{100}}, \bibinfo{pages}{693--700}
  (\bibinfo{year}{2011}).

\bibitem{Wright1999}
\bibinfo{author}{Wright, P.~E.} \& \bibinfo{author}{Dyson, H.~J.}
\newblock \bibinfo{title}{Intrinsically unstructured proteins: re-assessing the
  protein structure-function paradigm}.
\newblock \emph{\bibinfo{journal}{Journal of molecular biology}}
  \textbf{\bibinfo{volume}{293}}, \bibinfo{pages}{321--331}
  (\bibinfo{year}{1999}).

\bibitem{moruz2017peptide}
\bibinfo{author}{Moruz, L.} \& \bibinfo{author}{K{\"a}ll, L.}
\newblock \bibinfo{title}{Peptide retention time prediction}.
\newblock \emph{\bibinfo{journal}{Mass spectrometry reviews}}
  \textbf{\bibinfo{volume}{36}}, \bibinfo{pages}{615--623}
  (\bibinfo{year}{2017}).

\bibitem{Lijnzaad1997}
\bibinfo{author}{Lijnzaad, P.} \& \bibinfo{author}{Argos, P.}
\newblock \bibinfo{title}{Hydrophobic patches on protein subunit interfaces:
  characteristics and prediction}.
\newblock \emph{\bibinfo{journal}{Proteins: Structure, Function, and
  Bioinformatics}} \textbf{\bibinfo{volume}{28}}, \bibinfo{pages}{333--343}
  (\bibinfo{year}{1997}).

\bibitem{Bahadur2003}
\bibinfo{author}{Bahadur, R.~P.}, \bibinfo{author}{Chakrabarti, P.},
  \bibinfo{author}{Rodier, F.} \& \bibinfo{author}{Janin, J.}
\newblock \bibinfo{title}{Dissecting subunit interfaces in homodimeric
  proteins}.
\newblock \emph{\bibinfo{journal}{Proteins: Structure, Function, and
  Bioinformatics}} \textbf{\bibinfo{volume}{53}}, \bibinfo{pages}{708--719}
  (\bibinfo{year}{2003}).

\bibitem{Huang2000}
\bibinfo{author}{Huang, D.~M.} \& \bibinfo{author}{Chandler, D.}
\newblock \bibinfo{title}{Temperature and length scale dependence of
  hydrophobic effects and their possible implications for protein folding}.
\newblock \emph{\bibinfo{journal}{Proceedings of the National Academy of
  Sciences}} \textbf{\bibinfo{volume}{97}}, \bibinfo{pages}{8324--8327}
  (\bibinfo{year}{2000}).
\newblock \urlprefix\url{https://www.pnas.org/content/97/15/8324}.
\newblock \eprint{https://www.pnas.org/content/97/15/8324.full.pdf}.

\bibitem{Larsen1998}
\bibinfo{author}{Larsen, T.~A.}, \bibinfo{author}{Olson, A.~J.} \&
  \bibinfo{author}{Goodsell, D.~S.}
\newblock \bibinfo{title}{Morphology of protein–protein interfaces}.
\newblock \emph{\bibinfo{journal}{Structure}} \textbf{\bibinfo{volume}{6}},
  \bibinfo{pages}{421--427} (\bibinfo{year}{1998}).

\bibitem{dobson2004principles}
\bibinfo{author}{Dobson, C.~M.}
\newblock \bibinfo{title}{Principles of protein folding, misfolding and
  aggregation}.
\newblock In \emph{\bibinfo{booktitle}{Seminars in cell \& developmental
  biology}}, vol.~\bibinfo{volume}{15}, \bibinfo{pages}{3--16}
  (\bibinfo{organization}{Elsevier}, \bibinfo{year}{2004}).

\bibitem{Gomez1995}
\bibinfo{author}{Gomez, J.}, \bibinfo{author}{Hilser, V.~J.},
  \bibinfo{author}{Xie, D.} \& \bibinfo{author}{Freire, E.}
\newblock \bibinfo{title}{The heat capacity of proteins}.
\newblock \emph{\bibinfo{journal}{Proteins: Structure, Function, and
  Bioinformatics}} \textbf{\bibinfo{volume}{22}}, \bibinfo{pages}{404--412}
  (\bibinfo{year}{1995}).

\bibitem{Dijk2016}
\bibinfo{author}{Dijk, E.~V.}, \bibinfo{author}{Varilly, P.},
  \bibinfo{author}{Knowles, T. P.~J.}, \bibinfo{author}{Frenkel, D.} \&
  \bibinfo{author}{Abeln, S.}
\newblock \bibinfo{title}{Consistent treatment of hydrophobicity in protein
  lattice models accounts for cold denaturation}.
\newblock \emph{\bibinfo{journal}{Physical review letters}}
  \textbf{\bibinfo{volume}{116}}, \bibinfo{pages}{078101}
  (\bibinfo{year}{2016}).

\bibitem{kabsch1983dictionary}
\bibinfo{author}{Kabsch, W.} \& \bibinfo{author}{Sander, C.}
\newblock \bibinfo{title}{Dictionary of protein secondary structure: pattern
  recognition of hydrogen-bonded and geometrical features}.
\newblock \emph{\bibinfo{journal}{Biopolymers: Original Research on
  Biomolecules}} \textbf{\bibinfo{volume}{22}}, \bibinfo{pages}{2577--2637}
  (\bibinfo{year}{1983}).

\bibitem{Lijnzaad1996}
\bibinfo{author}{Lijnzaad, P.}, \bibinfo{author}{Berendsen, H. J.~C.} \&
  \bibinfo{author}{Argos, P.}
\newblock \bibinfo{title}{A method for detecting hydrophobic patches on protein
  surfaces}.
\newblock \emph{\bibinfo{journal}{Proteins: Structure, Function, and
  Bioinformatics}} \textbf{\bibinfo{volume}{26}}, \bibinfo{pages}{192--203}
  (\bibinfo{year}{1996}).

\bibitem{Garg2005}
\bibinfo{author}{Garg, A.}, \bibinfo{author}{Kaur, H.} \&
  \bibinfo{author}{Raghava, G. P.~S.}
\newblock \bibinfo{title}{Real value prediction of solvent accessibility in
  proteins using multiple sequence alignment and secondary structure}.
\newblock \emph{\bibinfo{journal}{Proteins: Structure, Function, and
  Bioinformatics}} \textbf{\bibinfo{volume}{61}}, \bibinfo{pages}{318--324}
  (\bibinfo{year}{2005}).

\bibitem{Petersen2009}
\bibinfo{author}{Petersen, B.}, \bibinfo{author}{Petersen, T.~N.},
  \bibinfo{author}{Andersen, P.}, \bibinfo{author}{Nielsen, M.} \&
  \bibinfo{author}{Lundegaard, C.}
\newblock \bibinfo{title}{A generic method for assignment of reliability scores
  applied to solvent accessibility predictions}.
\newblock \emph{\bibinfo{journal}{BMC structural biology}}
  \textbf{\bibinfo{volume}{9}}, \bibinfo{pages}{51} (\bibinfo{year}{2009}).

\bibitem{Joo2012}
\bibinfo{author}{Joo, K.}, \bibinfo{author}{Lee, S.~J.} \&
  \bibinfo{author}{Lee, J.}
\newblock \bibinfo{title}{Sann: solvent accessibility prediction of proteins by
  nearest neighbor method}.
\newblock \emph{\bibinfo{journal}{Proteins: Structure, Function, and
  Bioinformatics}} \textbf{\bibinfo{volume}{80}}, \bibinfo{pages}{1791--1797}
  (\bibinfo{year}{2012}).

\bibitem{Faraggi2012}
\bibinfo{author}{Faraggi, E.}, \bibinfo{author}{Zhang, T.},
  \bibinfo{author}{Yang, Y.}, \bibinfo{author}{Kurgan, L.} \&
  \bibinfo{author}{Zhou, Y.}
\newblock \bibinfo{title}{Spine x: improving protein secondary structure
  prediction by multistep learning coupled with prediction of solvent
  accessible surface area and backbone torsion angles}.
\newblock \emph{\bibinfo{journal}{Journal of computational chemistry}}
  \textbf{\bibinfo{volume}{33}}, \bibinfo{pages}{259--267}
  (\bibinfo{year}{2012}).

\bibitem{Klausen2019}
\bibinfo{author}{Klausen, M.~S.} \emph{et~al.}
\newblock \bibinfo{title}{Netsurfp‐2.0: Improved prediction of protein
  structural features by integrated deep learning}.
\newblock \emph{\bibinfo{journal}{Proteins: Structure, Function, and
  Bioinformatics}} \textbf{\bibinfo{volume}{87}}, \bibinfo{pages}{520--527}
  (\bibinfo{year}{2019}).

\bibitem{kyte1982}
\bibinfo{author}{Kyte, J.} \& \bibinfo{author}{Doolittle, R.~F.}
\newblock \bibinfo{title}{A simple method for displaying the hydropathic
  character of a protein}.
\newblock \emph{\bibinfo{journal}{Journal of Molecular Biology}}
  \textbf{\bibinfo{volume}{157}}, \bibinfo{pages}{105 -- 132}
  (\bibinfo{year}{1982}).
\newblock
  \urlprefix\url{http://www.sciencedirect.com/science/article/pii/0022283682905150}.

\bibitem{xu2020opus}
\bibinfo{author}{Xu, G.}, \bibinfo{author}{Wang, Q.} \& \bibinfo{author}{Ma,
  J.}
\newblock \bibinfo{title}{Opus-tass: a protein backbone torsion angles and
  secondary structure predictor based on ensemble neural networks}.
\newblock \emph{\bibinfo{journal}{Bioinformatics}}
  \textbf{\bibinfo{volume}{36}}, \bibinfo{pages}{5021--5026}
  (\bibinfo{year}{2020}).

\bibitem{fereshteh2020enhancing}
\bibinfo{author}{Fereshteh, M.} \emph{et~al.}
\newblock \bibinfo{title}{Enhancing protein backbone angle prediction by using
  simpler models of deep neural networks}.
\newblock \emph{\bibinfo{journal}{Scientific Reports (Nature Publisher Group)}}
  \textbf{\bibinfo{volume}{10}} (\bibinfo{year}{2020}).

\bibitem{kyte1982simple}
\bibinfo{author}{Kyte, J.} \& \bibinfo{author}{Doolittle, R.~F.}
\newblock \bibinfo{title}{A simple method for displaying the hydropathic
  character of a protein}.
\newblock \emph{\bibinfo{journal}{Journal of molecular biology}}
  \textbf{\bibinfo{volume}{157}}, \bibinfo{pages}{105--132}
  (\bibinfo{year}{1982}).

\bibitem{lobry1994hydrophobicity}
\bibinfo{author}{Lobry, J.} \& \bibinfo{author}{Gautier, C.}
\newblock \bibinfo{title}{Hydrophobicity, expressivity and aromaticity are the
  major trends of amino-acid usage in 999 escherichia coli chromosome-encoded
  genes}.
\newblock \emph{\bibinfo{journal}{Nucleic acids research}}
  \textbf{\bibinfo{volume}{22}}, \bibinfo{pages}{3174--3180}
  (\bibinfo{year}{1994}).

\bibitem{uniprot2019uniprot}
\bibinfo{author}{Consortium, U.}
\newblock \bibinfo{title}{Uniprot: a worldwide hub of protein knowledge}.
\newblock \emph{\bibinfo{journal}{Nucleic acids research}}
  \textbf{\bibinfo{volume}{47}}, \bibinfo{pages}{D506--D515}
  (\bibinfo{year}{2019}).

\bibitem{ponten2008human}
\bibinfo{author}{Pont{\'e}n, F.}, \bibinfo{author}{Jirstr{\"o}m, K.} \&
  \bibinfo{author}{Uhlen, M.}
\newblock \bibinfo{title}{The human protein atlas—a tool for pathology}.
\newblock \emph{\bibinfo{journal}{The Journal of Pathology: A Journal of the
  Pathological Society of Great Britain and Ireland}}
  \textbf{\bibinfo{volume}{216}}, \bibinfo{pages}{387--393}
  (\bibinfo{year}{2008}).

\bibitem{uhlen2015tissue}
\bibinfo{author}{Uhl{\'e}n, M.} \emph{et~al.}
\newblock \bibinfo{title}{Tissue-based map of the human proteome}.
\newblock \emph{\bibinfo{journal}{Science}} \textbf{\bibinfo{volume}{347}}
  (\bibinfo{year}{2015}).

\bibitem{lote1994principles}
\bibinfo{author}{Lote, C.~J.} \& \bibinfo{author}{Lote, C.~J.}
\newblock \emph{\bibinfo{title}{Principles of renal physiology}}.
\newblock \bibinfo{number}{QP211 L88 1994} (\bibinfo{publisher}{Springer},
  \bibinfo{year}{1994}).

\bibitem{kanehisa2000kegg}
\bibinfo{author}{Kanehisa, M.} \& \bibinfo{author}{Goto, S.}
\newblock \bibinfo{title}{Kegg: kyoto encyclopedia of genes and genomes}.
\newblock \emph{\bibinfo{journal}{Nucleic acids research}}
  \textbf{\bibinfo{volume}{28}}, \bibinfo{pages}{27--30}
  (\bibinfo{year}{2000}).

\bibitem{zheng2019deep}
\bibinfo{author}{Zheng, W.} \emph{et~al.}
\newblock \bibinfo{title}{Deep-learning contact-map guided protein structure
  prediction in casp13}.
\newblock \emph{\bibinfo{journal}{Proteins: Structure, Function, and
  Bioinformatics}} \textbf{\bibinfo{volume}{87}}, \bibinfo{pages}{1149--1164}
  (\bibinfo{year}{2019}).

\bibitem{li2019respre}
\bibinfo{author}{Li, Y.}, \bibinfo{author}{Hu, J.}, \bibinfo{author}{Zhang,
  C.}, \bibinfo{author}{Yu, D.-J.} \& \bibinfo{author}{Zhang, Y.}
\newblock \bibinfo{title}{Respre: high-accuracy protein contact prediction by
  coupling precision matrix with deep residual neural networks}.
\newblock \emph{\bibinfo{journal}{Bioinformatics}}
  \textbf{\bibinfo{volume}{35}}, \bibinfo{pages}{4647--4655}
  (\bibinfo{year}{2019}).

\bibitem{senior2020improved}
\bibinfo{author}{Senior, A.~W.} \emph{et~al.}
\newblock \bibinfo{title}{Improved protein structure prediction using
  potentials from deep learning}.
\newblock \emph{\bibinfo{journal}{Nature}} \textbf{\bibinfo{volume}{577}},
  \bibinfo{pages}{706--710} (\bibinfo{year}{2020}).

\bibitem{tartaglia2009}
\bibinfo{author}{Tartaglia, G.~G.}, \bibinfo{author}{Pechmann, S.},
  \bibinfo{author}{Dobson, C.~M.} \& \bibinfo{author}{Vendruscolo, M.}
\newblock \bibinfo{title}{A relationship between mrna expression levels and
  protein solubility in e. coli}.
\newblock \emph{\bibinfo{journal}{Journal of molecular biology}}
  \textbf{\bibinfo{volume}{388}}, \bibinfo{pages}{381--389}
  (\bibinfo{year}{2009}).

\bibitem{kramer2012toward}
\bibinfo{author}{Kramer, R.~M.}, \bibinfo{author}{Shende, V.~R.},
  \bibinfo{author}{Motl, N.}, \bibinfo{author}{Pace, C.~N.} \&
  \bibinfo{author}{Scholtz, J.~M.}
\newblock \bibinfo{title}{Toward a molecular understanding of protein
  solubility: increased negative surface charge correlates with increased
  solubility}.
\newblock \emph{\bibinfo{journal}{Biophysical journal}}
  \textbf{\bibinfo{volume}{102}}, \bibinfo{pages}{1907--1915}
  (\bibinfo{year}{2012}).

\bibitem{sytnyk2017neural}
\bibinfo{author}{Sytnyk, V.}, \bibinfo{author}{Leshchyns’ka, I.} \&
  \bibinfo{author}{Schachner, M.}
\newblock \bibinfo{title}{Neural cell adhesion molecules of the immunoglobulin
  superfamily regulate synapse formation, maintenance, and function}.
\newblock \emph{\bibinfo{journal}{Trends in neurosciences}}
  \textbf{\bibinfo{volume}{40}}, \bibinfo{pages}{295--308}
  (\bibinfo{year}{2017}).

\bibitem{stadelmann2019myelin}
\bibinfo{author}{Stadelmann, C.}, \bibinfo{author}{Timmler, S.},
  \bibinfo{author}{Barrantes-Freer, A.} \& \bibinfo{author}{Simons, M.}
\newblock \bibinfo{title}{Myelin in the central nervous system: structure,
  function, and pathology}.
\newblock \emph{\bibinfo{journal}{Physiological reviews}}
  \textbf{\bibinfo{volume}{99}}, \bibinfo{pages}{1381--1431}
  (\bibinfo{year}{2019}).

\bibitem{Cock2009}
\bibinfo{author}{Cock, P. J.~A.} \emph{et~al.}
\newblock \bibinfo{title}{{Biopython: freely available Python tools for
  computational molecular biology and bioinformatics}}.
\newblock \emph{\bibinfo{journal}{Bioinformatics}}
  \textbf{\bibinfo{volume}{25}}, \bibinfo{pages}{1422--1423}
  (\bibinfo{year}{2009}).
\newblock \urlprefix\url{https://doi.org/10.1093/bioinformatics/btp163}.
\newblock
  \eprint{https://academic.oup.com/bioinformatics/article-pdf/25/11/1422/944180/btp163.pdf}.

\bibitem{sanner1996reduced}
\bibinfo{author}{{Sanner}, M.~F.}, \bibinfo{author}{{Olson}, A.~J.} \&
  \bibinfo{author}{{Spehner}, J.-C.}
\newblock \bibinfo{title}{Reduced surface: An efficient way to compute
  molecular surfaces}.
\newblock \emph{\bibinfo{journal}{Biopolymers}} \textbf{\bibinfo{volume}{38}},
  \bibinfo{pages}{305--320} (\bibinfo{year}{1996}).

\bibitem{Bentley1975}
\bibinfo{author}{Bentley, J.~L.}
\newblock \bibinfo{title}{Multidimensional binary search trees used for
  associative searching}.
\newblock \emph{\bibinfo{journal}{Commun. ACM}} \textbf{\bibinfo{volume}{18}},
  \bibinfo{pages}{509–517} (\bibinfo{year}{1975}).
\newblock \urlprefix\url{https://doi.org/10.1145/361002.361007}.

\bibitem{wang2003pisces}
\bibinfo{author}{Wang, G.} \& \bibinfo{author}{Dunbrack~Jr, R.~L.}
\newblock \bibinfo{title}{Pisces: a protein sequence culling server}.
\newblock \emph{\bibinfo{journal}{Bioinformatics}}
  \textbf{\bibinfo{volume}{19}}, \bibinfo{pages}{1589--1591}
  (\bibinfo{year}{2003}).

\bibitem{sonnhammer1998hidden}
\bibinfo{author}{Sonnhammer, E.~L.}, \bibinfo{author}{Von~Heijne, G.},
  \bibinfo{author}{Krogh, A.} \emph{et~al.}
\newblock \bibinfo{title}{A hidden markov model for predicting transmembrane
  helices in protein sequences.}
\newblock In \emph{\bibinfo{booktitle}{Ismb}}, vol.~\bibinfo{volume}{6},
  \bibinfo{pages}{175--182} (\bibinfo{year}{1998}).

\bibitem{moller2001evaluation}
\bibinfo{author}{M{\"o}ller, S.}, \bibinfo{author}{Croning, M.~D.} \&
  \bibinfo{author}{Apweiler, R.}
\newblock \bibinfo{title}{Evaluation of methods for the prediction of membrane
  spanning regions}.
\newblock \emph{\bibinfo{journal}{Bioinformatics}}
  \textbf{\bibinfo{volume}{17}}, \bibinfo{pages}{646--653}
  (\bibinfo{year}{2001}).

\bibitem{Kuhn2008}
\bibinfo{author}{Kuhn, M.}
\newblock \bibinfo{title}{Building predictive models in r using the caret
  package}.
\newblock \emph{\bibinfo{journal}{Journal of Statistical Software, Articles}}
  \textbf{\bibinfo{volume}{28}}, \bibinfo{pages}{1--26} (\bibinfo{year}{2008}).
\newblock \urlprefix\url{https://www.jstatsoft.org/v028/i05}.

\bibitem{chen2016}
\bibinfo{author}{Chen, T.} \& \bibinfo{author}{Guestrin, C.}
\newblock \bibinfo{title}{Xgboost: A scalable tree boosting system}.
\newblock In \emph{\bibinfo{booktitle}{Proceedings of the 22nd ACM SIGKDD
  International Conference on Knowledge Discovery and Data Mining}}, KDD '16,
  \bibinfo{pages}{785–794} (\bibinfo{publisher}{Association for Computing
  Machinery}, \bibinfo{address}{New York, NY, USA}, \bibinfo{year}{2016}).
\newblock \urlprefix\url{https://doi.org/10.1145/2939672.2939785}.

\bibitem{Zemla2001}
\bibinfo{author}{Zemla, A.}, \bibinfo{author}{Venclovas, A.},
  \bibinfo{author}{Moult, J.} \& \bibinfo{author}{Fidelis, K.}
\newblock \bibinfo{title}{Processing and evaluation of predictions in casp4}.
\newblock \emph{\bibinfo{journal}{Proteins: Structure, Function, and
  Bioinformatics}} \textbf{\bibinfo{volume}{45}}, \bibinfo{pages}{13--21}
  (\bibinfo{year}{2001}).
\newblock
  \urlprefix\url{https://onlinelibrary.wiley.com/doi/abs/10.1002/prot.10052}.
\newblock \eprint{https://onlinelibrary.wiley.com/doi/pdf/10.1002/prot.10052}.

\bibitem{subramanian2005gene}
\bibinfo{author}{Subramanian, A.} \emph{et~al.}
\newblock \bibinfo{title}{Gene set enrichment analysis: a knowledge-based
  approach for interpreting genome-wide expression profiles}.
\newblock \emph{\bibinfo{journal}{Proceedings of the National Academy of
  Sciences}} \textbf{\bibinfo{volume}{102}}, \bibinfo{pages}{15545--15550}
  (\bibinfo{year}{2005}).

\bibitem{mootha2003pgc}
\bibinfo{author}{Mootha, V.~K.} \emph{et~al.}
\newblock \bibinfo{title}{Pgc-1$\alpha$-responsive genes involved in oxidative
  phosphorylation are coordinately downregulated in human diabetes}.
\newblock \emph{\bibinfo{journal}{Nature genetics}}
  \textbf{\bibinfo{volume}{34}}, \bibinfo{pages}{267--273}
  (\bibinfo{year}{2003}).

\bibitem{higurashi2009pisite}
\bibinfo{author}{Higurashi, M.}, \bibinfo{author}{Ishida, T.} \&
  \bibinfo{author}{Kinoshita, K.}
\newblock \bibinfo{title}{Pisite: a database of protein interaction sites using
  multiple binding states in the pdb}.
\newblock \emph{\bibinfo{journal}{Nucleic acids research}}
  \textbf{\bibinfo{volume}{37}}, \bibinfo{pages}{D360--D364}
  (\bibinfo{year}{2009}).

\bibitem{Janin1979}
\bibinfo{author}{Janin, J.}
\newblock \bibinfo{title}{{Surface and inside volumes in globular proteins}}.
\newblock \emph{\bibinfo{journal}{Nature}} \textbf{\bibinfo{volume}{277}},
  \bibinfo{pages}{491--492} (\bibinfo{year}{1979}).
\newblock \urlprefix\url{http://dx.doi.org/10.1038/277491a0}.

\bibitem{Chothia1976}
\bibinfo{author}{Chothia, C.}
\newblock \bibinfo{title}{{The nature of the accessible and buried surfaces in
  proteins}}.
\newblock \emph{\bibinfo{journal}{Journal of Molecular Biology}}
  \textbf{\bibinfo{volume}{105}}, \bibinfo{pages}{1--12}
  (\bibinfo{year}{1976}).
\newblock
  \urlprefix\url{http://www.sciencedirect.com/science/article/pii/0022283676901911}.

\end{thebibliography}

\clearpage
\newpage

\appendix

\renewcommand{\thefigure}{S\arabic{figure}}
\renewcommand{\theequation}{S\arabic{equation}}
\renewcommand{\thetable}{S\arabic{table}}
\setcounter{figure}{0}
\setcounter{equation}{0}
\setcounter{table}{0}

\section{Supporting Information}

\subsection{Centering and scaling of the THSA, RHSA and LHP distributions for GSEA}
\label{sec:centering}
\begin{figure}[h]
    \centering
    \includegraphics[width=.9\linewidth]{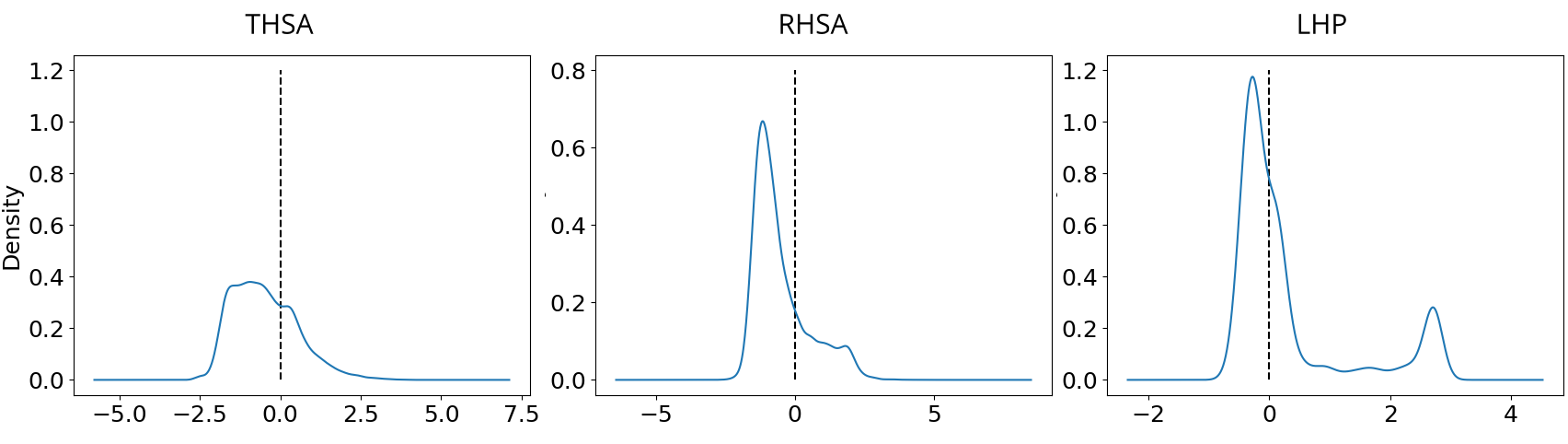}
    \caption{\textbf{Distributions of the centered and scaled values of the THSA, RHSA and LHP used for GSEA analysis}. The dotted lines indicate the zero positions. Values for centering were chosen such that 0 falls in between two parts of a bimodal distribution or between the main bulk and the tail of the distribution.}
    \label{fig:centering}
\end{figure}

THSA, RHSA, and LHP values were centered and scaled prior GSEA analysis. To get biologically meaningful values, the values for centering were chosen such that 0 fell in between two parts of a bimodal distribution or between the main bulk and the tail of the distribution~\ref{fig:centering}:

\begin{equation}
    x_{i_{new}} = \frac{x_{i_{old}} - center}{\sigma}
    \label{eq:centering}
\end{equation}

\noindent where $x_{i_{new}}$ is the centered and scaled value for protein $i$, $x_{i_{old}}$ is the original value (THSA, RHSA or LHP) of protein $i$ and center is the zero position chosen based on the original distributions: 8106 \AA~for the THSA, 0.35 for the RHSA and 1656 \AA for the LHP, see Figure \ref{fig:centering}.


\subsection{Overlap between protein-protein interaction sites and the LHP}
\begin{figure}
    \centering
    \includegraphics[width=0.95\linewidth]{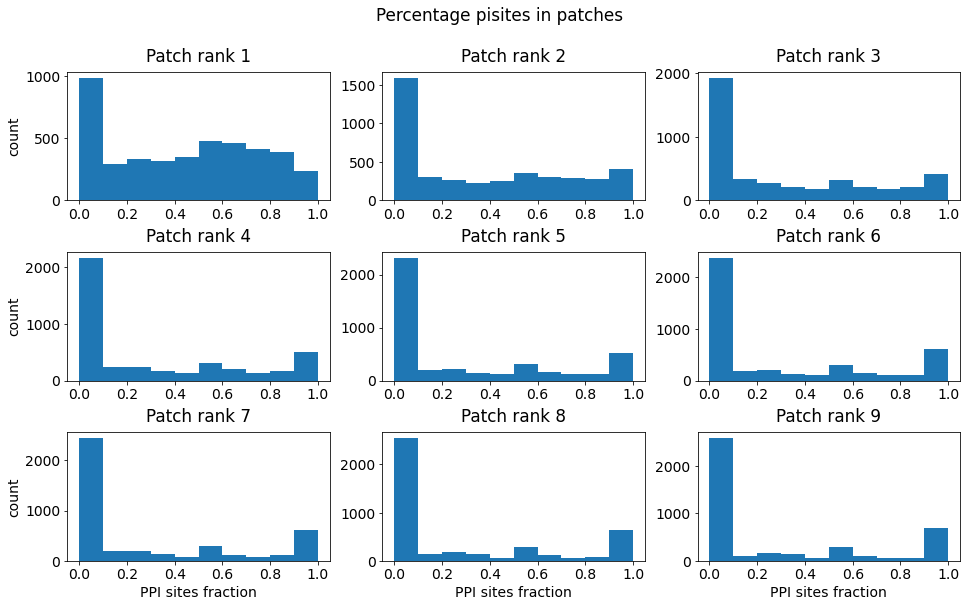}
    \caption{\textbf{Distribution of interaction sites over the nine largest hydrophobic patches per protein}. For each patch on a protein we calculated which fraction of the hydrophobic patch on that protein overlaps with protein-protein interaction sites for the nine largest hydrophobic patches. The three largest patches in each protein have a significantly larger overlap with the PPIs than the other ones.}
    \label{fig:pisite}
\end{figure}

The data set with the information about protein binding sites was obtained via the PiSITE database. Both interaction information from a single PDB complex and interaction information between multiple PDBs are stored \citep{higurashi2009pisite}. Only PiSITE information of the proteins from the original 14,602 chains dataset was included. The proteins without interaction sites and transmembrane proteins were filtered, which resulted in the data set of 4,255 entries with information about protein interaction sites. Figure~\ref{fig:pisite} shows that the patches with a higher rank (i.e. the larger patches of each protein) have a larger overlap with PPI sites. The three largest patches per protein have a significantly larger overlap with PPI sites than would be expected, as determined using a Wilcoxon rank sum test (Table~\ref{tab:pisite}).

\subsection{Benchmark of older methods}
\begin{figure}[h]
    \centering
    \includegraphics[width=.4\linewidth]{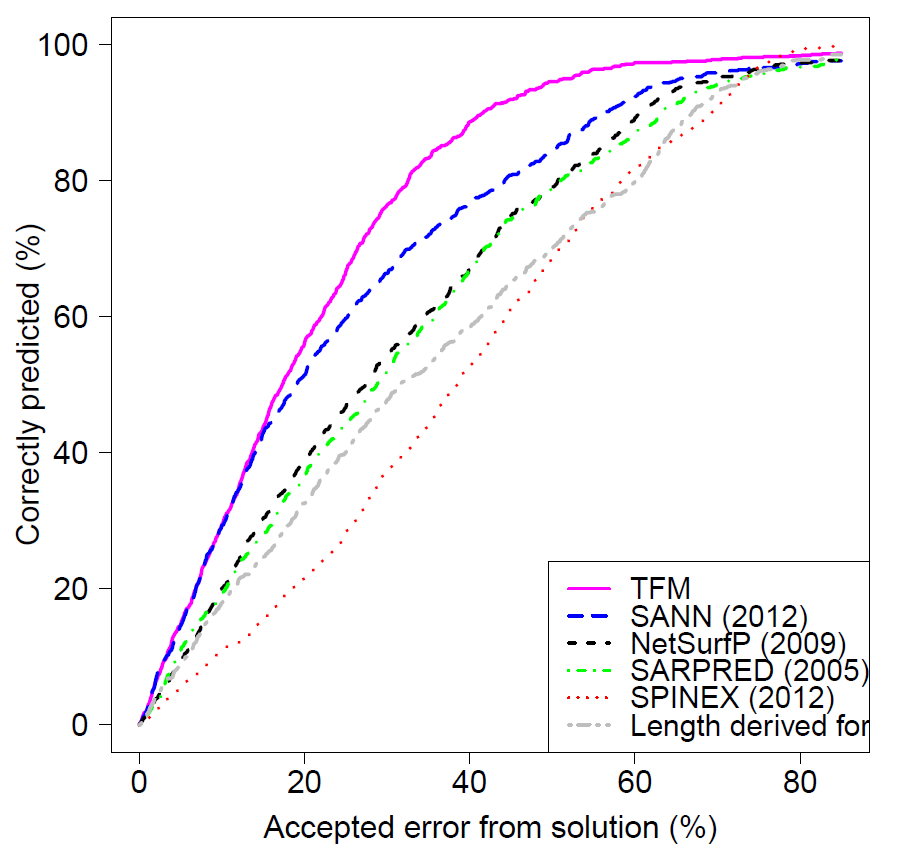}
    \caption{\textbf{Benchmark of SANN, NetsurfP, SARPRED, SPINEX and LBM for hydrophobic surface area predictions}. The figure shows that the TFM outperforms the other methods, indicating that the length and hydrophobicity of the sequence are very important features for predicting the surface area.}
    \label{fig:benchmark_old}
\end{figure}

Previous (unpublished) results of a benchmark of the TFM against SANN \citep{Joo2012}, NetsurfP (NOT the 2.0 version) \citep{Petersen2009}, SARPRED \citep{Garg2005}, SPINEX \citep{Faraggi2012} and a simple length-based reference model for predicting the THSA, as shown in Figure~\ref{fig:benchmark_old}.

SANN, NetsurfP, SARPRED and SPINEX were run using their default setting. Each of these methods predicts the surface area per residue. We summed over the predictions of all the hydrophobic residues to obtain the THSA. 

For the purpose of model comparison, we also developed a length-based reference model. This simple model provides an HSA estimate based on the length of the protein sequence. The idea of approximating proteins as a sphere to predict the ASA of the whole protein was first introduced in \citep{Janin1979}. The ratio between hydrophilic and hydrophobic residues on the surface has previously been observed in \citep{Chothia1976}: for proteins with a high molecular weight the ratio of hydrophobic residues can be well approximated for globular proteins based on the length of the protein sequence alone.

The reference model uses the sequence length of a protein ($L$) multiplied with a constant ($k_1$) and to the power of a constant ($k_2$) to predict the HSA:

\begin{equation}
	\label{eq:sphereModel}
	ASA = k_1 \cdot L^{k_2}
\end{equation} 

Note that in case of a perfect sphere, we would have: 

\begin{equation}
	\text{surface area} = 4\pi\left(\frac{3V}{4\pi}\right)^\frac{2}{3}
\end{equation}

Using the latter equation the total ASA could be approximated by assuming the sequence length ($L$) scales linearly with the volume ($V$). However, since proteins are not perfect spheres and only a fraction of the surface is covered by hydrophobic groups, we instead generate the baseline model by fitting the constants $k_1$ and $k_2$ to the training set, minimising the sum of squares between the predicted and observed HSA. In this simple model, we effectively assume that the fraction of hydrophobic amino acids on the surface with respect to the length is constant. 

Surprisingly, the TFM outperforms all other methods including NetsurfP \citep{Petersen2009}, which incorporates more information (evolutionary profiles) and has a more complicated architecture (neural network).


\subsection{Supplementary tables}

\begin{table}[hb]
    \centering
    \caption{P-values of the overlap of protein-protein interaction sites with the largest hydrophobic patches, calculated using a Wilcoxon rank sum test. Rank 1 indicates the largest patch of each protein, rank 2 the second largest patch, etc.}
    \begin{tabular}{@{}ll@{}}\toprule Rank & P-value \\\midrule
        1 & $2.59 \cdot 10^{-266}$ \\
        2 & $5.66 \cdot 10^{-24}$ \\
        3 & $3.41 \cdot 10^{-6}$ \\
        4 & $0.72$ \\
        5 & $0.66$ \\\bottomrule
    \end{tabular}
    \label{tab:pisite}
\end{table}

\clearpage
\onecolumn
\subsection{Supplementary figures}

\begin{figure}[h]
    \centering
    \includegraphics[width=\linewidth]{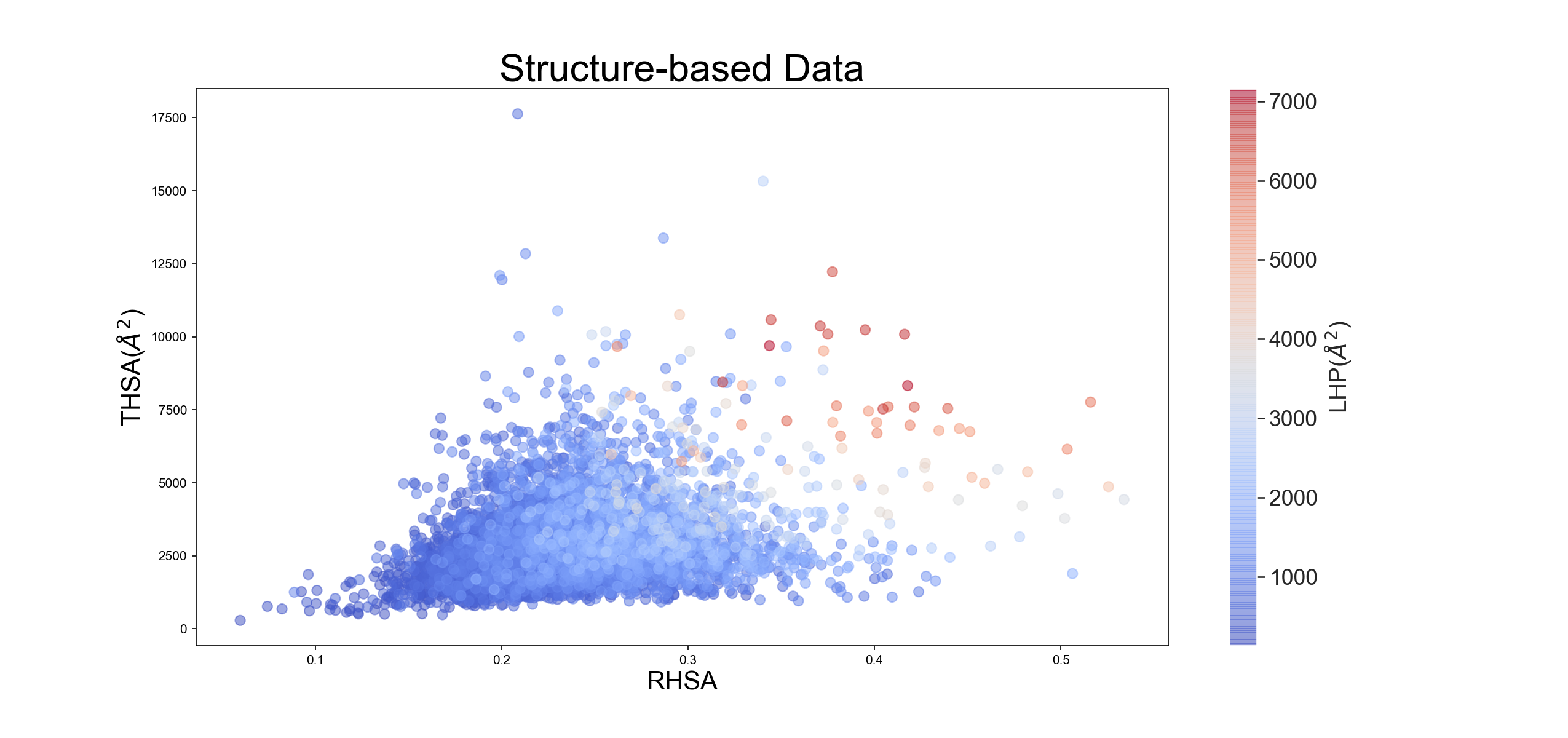}
    \caption{\textbf{Scatter plot showing the distribution of proteins in the structure-based data set based on THSA and RHSA values.} LHP values are colour-coded.}
    \label{fig:Structure_based_scatterplot}
\end{figure}

\begin{figure}
    \centering
    \includegraphics[width=\linewidth]{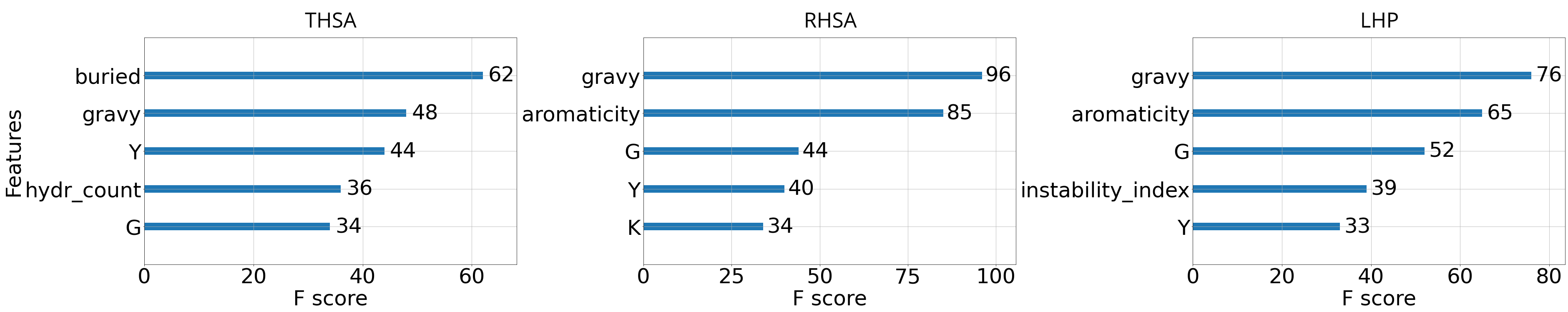}
    \caption{\textbf{Feature importance of the GFM for the THSA, RHSA and LHP predictions.} The five most important features in the GFM for each of the measures of surface hydrohpobicity were extracted using the XGBoost Python package \citep{chen2016}. The letters represent amino acids. A higher F score indicates a higher importance. One can see that in all cases hydrophobicity (hydr\_count, gravy, aromaticity~\citep{kyte1982, lobry1994hydrophobicity}) is important for the predictions.}
    \label{fig:feature_importance}
\end{figure}

\begin{figure}
  \centerline{\includegraphics[width=\linewidth]{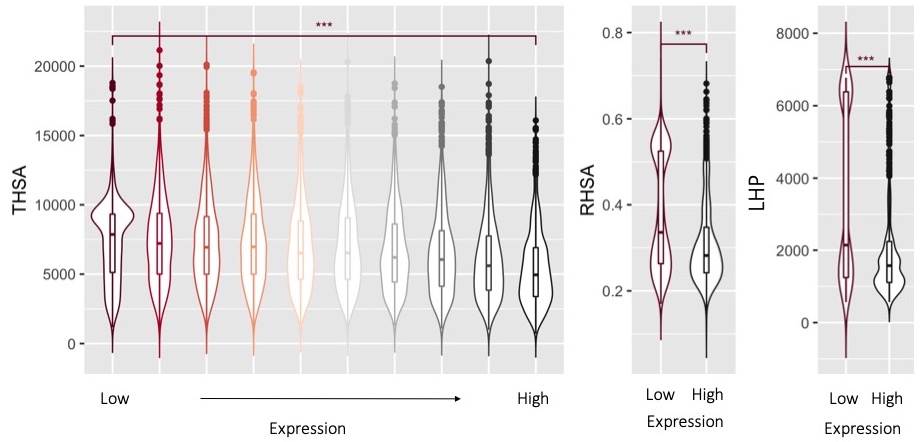}}
  \caption{\textbf{The relationship between normalised expression (NX) and THSA, RHSA and LHP values, respectively.} The median of NX values per gene was calculated across all tissues in which a particular gene occurs and proteins were grouped (n=3) equally from low to high expression (low, medium and high NX values). The differences between the groups of proteins with the lowest and the highest NX values was calculated using Wilcoxon signed-rank test. The level of significance is annotated with two (p-value: $<$ 0.01) and three asteriks (p-value: $<$ 2.22e-16).}
  \label{fig:expression_median}
\end{figure}

\begin{figure}[tbh!]
    \centering
    \includegraphics[width=0.5\linewidth]{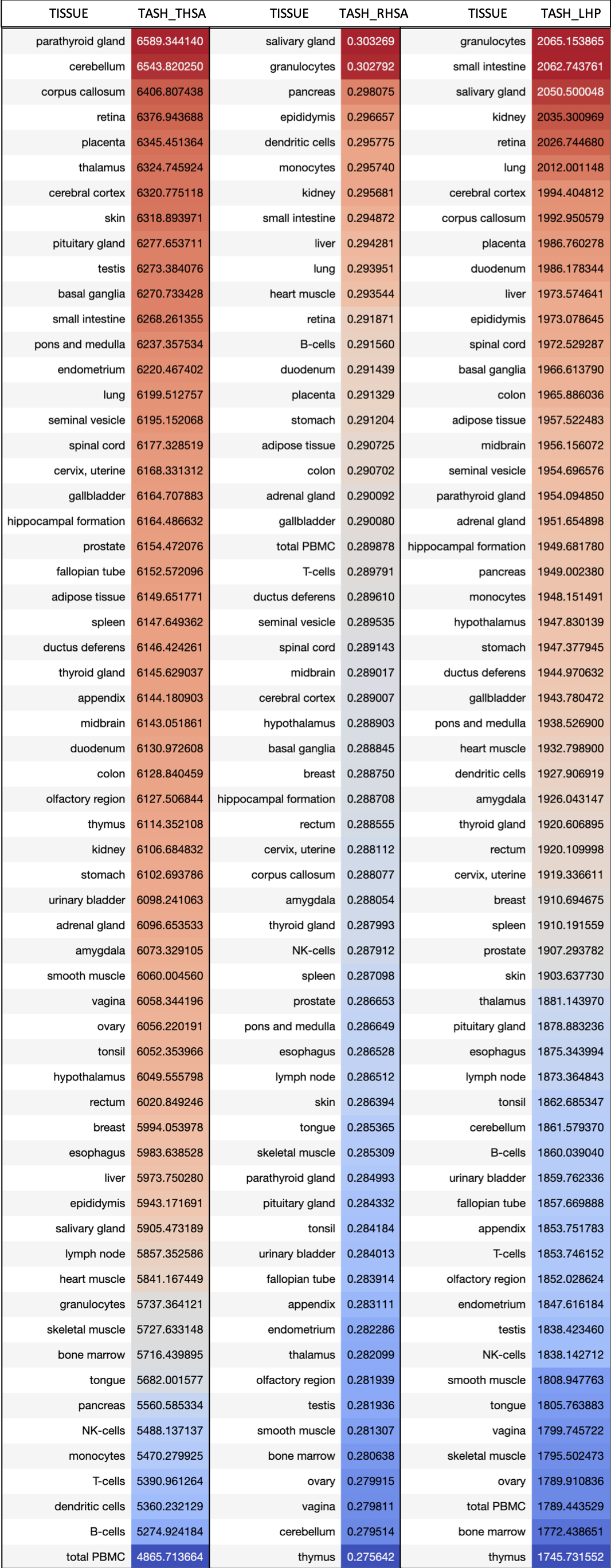}
    \caption{\textbf{Tissue-specific average surface hydrophobicity calculated (Equation~\ref{eq:TASH}) for different hydrophobic measures.} Each column is independently sorted and colour-coded based on TASH values.}
    \label{fig:TASH}
\end{figure}

\begin{figure}[tbh!]
    \centering
    \includegraphics[width=0.8\linewidth]{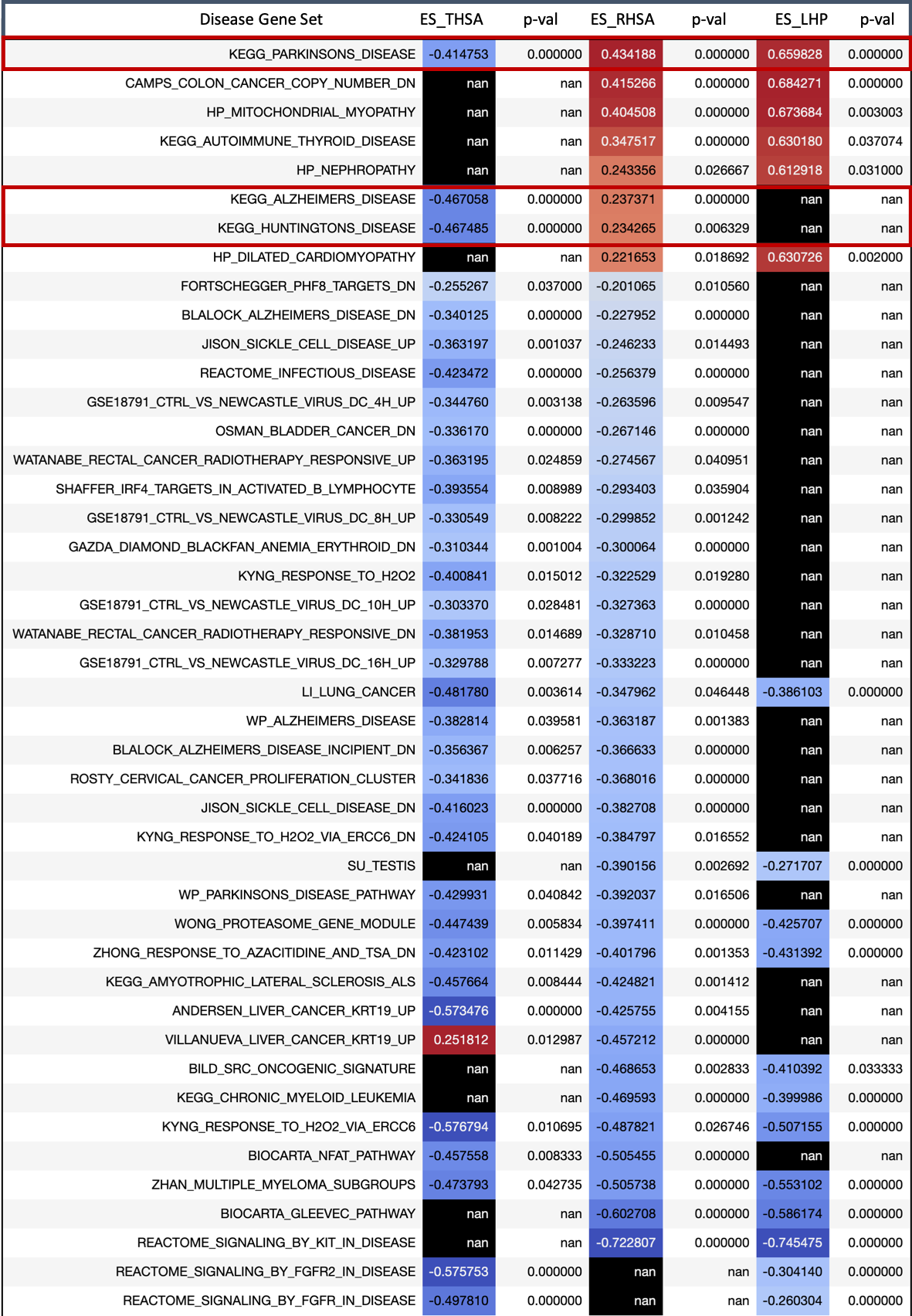}
    \caption{\textbf{Pre-ranked GSEA enrichment statistics in disease gene sets (n=375).} The values were central-scaled prior to the GSEA analysis. The enrichment score (ES) is the maximum deviation from zero showing the degree to which the gene set is over-represented at the top (positive ES score) or bottom (negative ES score) of the entire ranked list of genes. Disease gene sets with the nominal p-value $<$ 0.05 and ES $<$ -0.2 (negative enrichment) and ES > 0.2 (positive enrichment) were selected and kept only those that were significant in at least two hydrophobic measures. KEGG neurodegenerative pathways are highlighted with the red squares. 'Nan' value indicates that an ES score was either between -0.2;0.2 or insignificant (p-value $>$ 0.05).}
    \label{fig:GSEA_Diseases}
\end{figure}

\begin{figure}[h]
    \centering
    \includegraphics[width=.6\linewidth]{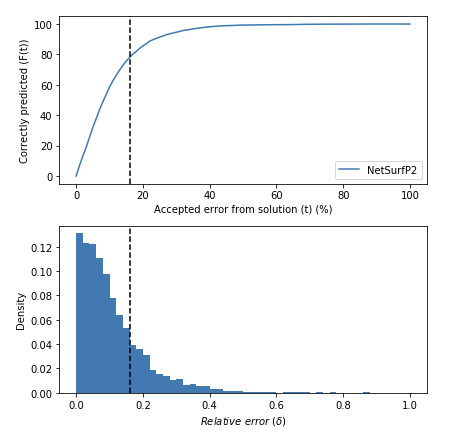}
    \caption{\textbf{An example of the relative threshold-based evaluation metric.} In this case, a threshold range from 0 to 100 percent absolute error is used to derive at the curve in the top panel (Equations~\ref{eq:THSA_error}, ~\ref{eq:RHSA_error}, ~\ref{eq:LHP_error}). For each threshold (exemplified by the black dashed horizontal line), the fraction of correctly predicted proteins within this threshold is calculated (Equation~\ref{eq:fraction_correct}). This fraction is the density to the left of the threshold.}
    \label{fig:error_example}
\end{figure}

\end{document}